\documentclass{article}

\usepackage{arxiv}

\usepackage{graphicx}
\usepackage{amsmath,amssymb,amsfonts,amsthm}
\usepackage{enumitem}
\usepackage{xcolor}
\usepackage{picins}
\usepackage{hyperref}

\newenvironment{pf}[1][PROOF.]%
  {\par\noindent\textbf{#1}\quad}
  {\par}

\theoremstyle{plain}
\if@secthm
  \newtheorem{thm}{Theorem}[section]
  \@addtoreset{thm}{section}
\else
  \newtheorem{thm}{Theorem}
\fi

\newtheorem{assum}[thm]{Assumption}

\theoremstyle{definition}

\newtheorem{rem}[thm]{Remark}

\title{KPCA for thrust vectoring systems exhibiting singular points}

\date{\today}

\author{ \href{https://orcid.org/0000-0002-0270-1199}{\includegraphics[scale=0.06]{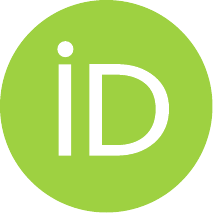}\hspace{1mm}Tam W.~Nguyen}
  \\
	Department of Electrical And Electronic Engineering\\
	University of Toyama\\
	Toyama 930 8555, Japan\\
	\texttt{nguyen@eng.u-toyama.ac.jp} \\
	\And
	\href{https://orcid.org/0000-0002-4986-2053}
 {\includegraphics[scale=0.06]{orcid.pdf}\hspace{1mm}Kyoungseok~Han} \\
	Department of Automotive Engineering \\
	Hanyang University \\
	Seoul, 04763, Republic of Korea\\
	\texttt{kyoungsh@hanyang.ac.kr} \\
	\AND
    \href{https://orcid.org/0000-0002-9767-4102}
	{\includegraphics[scale=0.06]{orcid.pdf}\hspace{1mm}Kenji~Hirata} \\
	Department of Electrical And Electronic Engineering \\
    University of Toyama \\
	Toyama 930 8555, Japan \\
	\texttt{hirata@eng.u-toyama.ac.jp} \\
}

\renewcommand{\shorttitle}{KPCA for thrust vectoring systems with singular points}

\begin{document}

\maketitle

\begin{abstract}                          
This paper considers a class of thrust vectoring systems, which are nonlinear, overactuated, and time-invariant.
We assume that the system is composed of two subsystems and there exist singular points around which the linearized system is uncontrollable.
Furthermore, we assume that the system is stabilizable through a two-level control allocation.
In this particular setting, we cannot do much with the linearized system, and a direct nonlinear control approach must be used to analyze the system stability.
Under adequate assumptions and a suitable nonlinear continuous control-allocation law, we can prove uniform asymptotic convergence of the points of equilibrium using Lyapunov input-to-state stability and the small gain theorem.
This control allocation, however, requires the design of an allocated mapping and introduces two exogenous inputs.
In particular, the closed-loop system is cascaded, and the output of one subsystem is the disturbance of the other, and vice versa.
In general, it is difficult to find a closed-form solution for the allocated mapping; it needs to satisfy restrictive conditions, among which Lipschitz continuity to ensure that the disturbances eventually vanish.
Additionally, this mapping is in general nontrivial and non-unique.
In this paper, we propose a new kernel-based predictive control allocation to substitute the need for designing an analytic mapping, and assess if it can produce a meaningful mapping ``on-the-fly" by solving online an optimization problem.
The simulations include three examples, which are the manipulation of an object through an unmanned aerial vehicle in two and three dimensions, and the control of a surface vessel actuated by two azimuthal thrusters.
\end{abstract}







\newpage 
\section{Introduction}

In the last decade, we have seen an increase of the use of unmanned aerial vehicles (UAVs) in a range of military 
\cite{7397888} and civilian applications 
\cite{8682048}, \cite{doi:10.1142/S2301385020500090}.
As those flying objects are getting increasingly more capable, we have thought to operate them as robots to physically interact with the environment \cite{5513152}, \cite{Maza2010}.
In these particular applications, the UAVs can be seen as effective orientable thrusters \cite{NGUYEN2021109586}, which can produce a force in all directions.
For example, we can control the UAV attitude through its four propellers and modify the thrust direction to transport \cite{SHIRANI2019158} and manipulate objects \cite{NGUYEN2019384}.
In such applications, as a general statement, we can say that the primary control objective is to create the required forces and torques to achieve the overall desired motion.
However, in numerous aerospace \cite{10.1007/978-3-030-33950-0_8},
\cite{6868215} and marine
\cite{https://doi.org/10.1002/rob.20369}, \cite{WITKOWSKA2018570} applications, the system is usually overactuated with a redundant set of actuators.
A possible solution is to follow a control-allocation strategy \cite{JOHANSEN20131087}.
It is based on a hierarchical motion-control approach and introduces one or more secondary objectives.
Note that adding more control objectives besides the primary objective is not a disadvantage per se, as it offers the designer extra degrees of freedom to minimize, for example, the power or fuel consumption of the system \cite{6315084}.
In the control-allocation framework, the primary control objective is achieved through a high-level motion control unit, which commands the overall forces and torques applied to the system.
A lower-level control-allocation unit is added to coordinate the different actuators such that they produce the desired forces and torques. 
This can be achieved by setting up the secondary objectives.

In this paper, we apply the control-allocation philosophy to a particular class of nonlinear, time-invariant, overactuated, and linearized-uncontrollable systems, which consist of two subsystems.
%
%
%
Similarly to \cite{4434782} and \cite{TJONNAS2007852}, we consider the actuator dynamics and thoroughly analyze the system stability through nonlinear control tools.
In particular, we show that we can stabilize the system by breaking down the overall control architecture into two separate control units for each subsystem.
Specifically, assuming that the other subsystem dynamics do not interfere, each control unit is designed such that the subsystem is stabilized.
Next, to achieve the primary control objective, we connect both subsystems through an allocated mapping.
This mapping is designed by properly formulating the secondary objective such that the closed-loop system guarantees that the desired forces and torques are generated.
This control architecture uses a cascaded structure and generates two disturbances entering each subsystem.
Under adequate assumptions, we can formally prove the system stability using Lyapunov input-to-state stability and the small gain theorem \cite{Khalil:1173048}.
However, as the system is overactuated, this allocated mapping is in general nontrivial and not unique.
Additionally, to guarantee that the disturbances eventually vanish and, thus, asymptotic stability, Lipschitz continuity of the mapping must be ensured.
Depending on the system complexity, finding such a closed-form mapping can be very hard.

The generalized (Moore-Penrose or pseudo) inverse \cite{johnson1985matrix} solution is the most commonly used method to compute the allocated mapping from the generalized force. This inverse, which is suggested in \cite{6315084} for control allocation, can either be performed on the matrix $B$ of the linearized system \cite{doi:10.2514/6.2002-6020}, \cite{doi:10.2514/3.21497}, \cite{doi:10.2514/3.21072} or on the Jacobian of the nonlinear system
\cite{5985959}, \cite{1386824}. Since overactuated systems possess infinite mappings and the generalized inverse always exists and is unique, this method is certainly appealing. However, our paper considers a special class of systems where the generalized inverse leads to oscillatory behaviors.
%
In fact, due to the system overactuation and the presence of scleronomic constraints, there exists a set of ``linearized-uncontrollable" points that must be stabilized. To ensure that all equilibrium points are asymptotically stable, the allocated mapping must be Lipschitz continuous.
Because the generalized-inverse method does not guarantee this condition to be true in general, applying it to the systems considered in this paper leads to oscillations.

For these reasons, this paper proposes a numerical solution to alleviate the difficulty of finding such an analytic Lipschitz-continuous mapping.
The main contribution of this paper is to propose the novel \emph{nonlinear kernel-based predictive control allocation} (KPCA).
KPCA introduces in the cost function a new term, which penalizes the deviation of the mapping from the kernel space.
By doing so, we are able to locally ``smooth out" the allocated mapping in the vicinity of the kernel space such that the system stability is preserved.
We investigate through three relevant numerical examples whether KPCA is able to produce a meaningful mapping ``on-the-fly" by solving online an optimization problem.

Note that a vaguely similar idea of combining nullspace techniques with quadratic programming (QP) but with linearization technique is suggested in \cite{9511202}. Although the authors tackle the similar problem of infinite mappings due to the system overactuation, they still use the generalized-inverse method to derive the mapping and the purpose of the nullspace is totally different from our paper. In particular, the authors first use the force-decomposition technique \cite{8485627}, which transforms the nonlinear allocation problem into a linear one. Next, to eliminate the linear approximation errors, the authors use the nullspace projection technique. However, the authors clearly state that they are unable to deal with the singularities, which are the core problem of this paper. In fact, \cite{9511202} avoids the singularities by enforcing that the angular changes are sufficiently small. There are two fundamental issues with this technique: i) since the mapping is derived from the generalized-inverse method, Lipschitz continuity of the allocated mapping is not guaranteed, and ii) any linearization technique would fail at the singular points. Therefore, it is necessary to directly tackle this problem through nonlinear control techniques.

%
%
Another contribution of this paper is the extension of past results \cite{doi:10.2514/1.G004356}, \cite{NGUYEN2019384}, \cite{NGUYEN2021109586} to more general systems other than aerospace.
As pointed out in \cite{JOHANSEN20131087}, cross-disciplinary research should be encouraged.
We thus attempt to give a more general formulation so that the proposed scheme can be modified and applied to other systems.
In particular, we demonstrate the effectiveness of KPCA by applying it to a surface vessel actuated by two azimuthal thrusters, where the analytic continuous stabilizing control law and the allocated mapping are a-priori unknown.

%

\section{Notation}

Depending on the variable dimension, ``$0$" should be taken as a scalar or a vector.
We denote by $\mathbb{N}$ the set of nonzero natural numbers, $\mathbb{N}_0$ the set of natural numbers with zero, $\mathbb{Z}$ the set of integers, $\mathbb{R}$ the set of real numbers, $\mathbb{R}^n$ the set of $n$-dimension real numbers, $\mathbb{R}^{n\times m}$ the set of $n$-by-$m$ real matrices, and $\mathbb{R}_{>0}$ and $\mathbb{R}_{\ge0}$ the sets of positive real numbers and semi-positive real numbers, respectively.
Likewise, we define $\mathbb{R}^{n\times m}_{>0}$ and $\mathbb{R}^{n\times m}_{\ge0}$ as the sets of real positive matrices and real semi-positive matrices, respectively.
The $SO(n)$ manifold and the set of $n$-by-$n$ skew matrices are defined by $SO(n)\triangleq \{\mathcal{R}\in\mathbb{R}^{n\times n}: \mathcal{R}\mathcal{R}^\mathrm{T} = I_n, \det(\mathcal{R})=1\}$ and $so(n)\triangleq\{\Omega\in\mathbb{R}^{n\times n}: \Omega^\mathrm{T} = -\Omega\},$ respectively, where $I_n$ is the $n$-by-$n$ identity matrix.
A rotation matrix $\mathcal{R}\in SO(3)$ can be parameterized \cite{math11122727} by the quaternion $\mathbf{q} \triangleq [q_0 \; q_v^\mathrm{T}]^\mathrm{T}\in\mathbb{H},$ where $q_0\in\mathbb{R}$ is the real part, $q_v\in\mathbb{R}^3$ is the imaginary part, $\|\mathbf{q}\|=1,$ and $\mathbb{H}$ is the Hilbert space.
The function $f$ is the application from $S$ to $T,$ that is, $f:S \to T:y=f(x),$ where $x\in S$ and $y\in T.$
For simplicity, we sometimes omit the arguments in $f(x_1, \ldots, x_n)$ and equivalently use $f(\cdot).$
The set of $n$-times differentiable functions is denoted by $C^n,$
the $n$-by-$n$ diagonal matrix is denoted by $D=\mathrm{diag}\{d_1,\ldots, d_i, \ldots,d_n\},$ where $d_i\in\mathbb{R}$ is the $i$-th diagonal entry,
the $\ell_2$ norm is denoted by $\|\cdot\|_2,$ and the $m$-by-$n$ matrix of zeros by $0_{m\times n}.$
A class $\kappa$ function is a strictly increasing, continuous function $\alpha:[0,a)\to[0,\infty),$ where $\alpha(0)=0.$ A class $\kappa\ell$ function is a continuous function $\beta:[0,a)\times[0,\infty)\to[0,\infty),$ where, for a fixed $s,$ the mapping $\beta(r,s)$ belongs to class $\kappa$ with respect to $r$ and, for each fixed $r,$ the mapping $\beta(r,s)$ is decreasing with respect to $s$ and $\beta(r,s)\to 0$ as $s\to\infty.$ We denote by $\sup_{x\in D} f(x)$ the supremum of $f$ in the domain $D.$

\section{A Class of Thrust Vectoring Systems with Singular Points}


Consider a system composed of Subsystem 1 and Subsystem 2, which are denoted by $\mathrm{S}_1$ and $\mathrm{S}_2,$ respectively.
Let $t\in\mathbb{R}_{\geq 0}$ be the continuous time, $x_1(t)\in\mathbb{R}^{n_1}$ and $x_2(t)\in\mathbb{R}^{n_2}$ be the states of $\mathrm{S}_1$ and $\mathrm{S}_2,$ respectively, and $u_1(t)\in\mathbb{R}^{m_1}$ and $u_2(t)\in\mathbb{R}^{m_2}$ be the inputs of $\mathrm{S}_1$ and $\mathrm{S}_2,$ respectively.
For notational simplicity, in this paper, we omit the time dependency of all the variables, that is, $x_1 \triangleq x_1(t),$ $x_2 \triangleq x_2(t),$ $u_1 \triangleq u_1(t),$ $u_2 \triangleq u_2(t),$ and so on without further notice. 

Consider the nonlinear time-invariant system
\begin{align}
    \mathrm{S}_1 \equiv \dot{x}_1 &= f(x_1, x_2, u_1), \label{eq:S1} \\
    \mathrm{S}_2 \equiv \dot{x}_2 &= g(x_2, u_2), \label{eq:S2}
\end{align}
where $f:\mathbb{R}^{n_1}\times\mathbb{R}^{n_2}\times\mathbb{R}^{m_1} \to \mathbb{R}^{n_1}$ and $g:\mathbb{R}^{n_2}\times\mathbb{R}^{m_2} \to \mathbb{R}^{n_2}$ are continuous and nonlinear.
%
%
From $(x_1, x_2, u_1),$ $\mathrm{S}_1$ generates the \emph{effective control}
\begin{align}\label{eq:Psi}
    \tilde{u}_1 &= \Psi(x_1, x_2, u_1),
\end{align}
where $\Psi: \mathbb{R}^{n_1 \times n_2 \times m_1}\to\mathbb{R}^{\tilde{m}_1}$ is nonlinear and continuous.
Consider the input constraint $u_{\mathrm{min}}\le u \le u_{\mathrm{max}},$ where $u\triangleq[u_1^\mathrm{T} \; u_2^\mathrm{T}]^\mathrm{T},$ $u_{\mathrm{min}} < u_{\mathrm{max}} \in \mathbb{R}^{m_1+m_2},$ and the inequalities are taken element-wise.
We assume the following.
\begin{assum}\label{ass:overactuated}
The system \eqref{eq:S1}-\eqref{eq:Psi} is overactuated, that is, $\dim(u_1) + \dim(x_2) = m_1 + n_2 > \dim(\tilde{u}_1) = \tilde{m}_1.$
\end{assum}
\begin{rem}
Since the system is overactuated, the solution to \eqref{eq:Psi} is in general not unique.
Note that, according to \eqref{eq:S2}, the dynamics of $x_2$ are directly influenced by $u_2.$
Thus, since $x_2$ is controlled by $u_2,$ we can interpret $x_2$ as a virtual control variable of $\mathrm{S}_1.$
\end{rem}
\begin{assum}\label{ass:kernel}
The kernel of $\Psi(\cdot),$ which is defined by
\begin{align}\label{eq:kernel}
\ker(\Psi(\cdot)) \triangleq \{(x_1, x_2, u_1): \Psi(x_1,x_2,u_1) = 0\},
\end{align}
is nonempty and satisfies the following condition. 
Let $\mathcal{K} \in C^0:\mathbb{R}^{n_1}\to\mathbb{R}^{n_2}.$ For all
$x_1 \in \mathbb{R}^{n_1},$ there exists $u_1^*\in\mathbb{R}^{m_1}$ such that $x_2^* = \mathcal{K}(x_1) \Rightarrow \Psi(x_1,x_2^*,u_1^*) = 0.
$
\end{assum}

%

\begin{assum}\label{ass:controllability}
There exist equilibrium points around which the linearized system of \eqref{eq:S1}, \eqref{eq:S2} is uncontrollable.
That is, defining $x \triangleq [x_1^\mathrm{T} \; x_2^\mathrm{T}]^\mathrm{T},$ $\bar{x}\in\mathrm{R}^{n_1+n_2},$ $\bar{u}\in\mathrm{R}^{m_1+m_2},$ $\tilde{x} \triangleq x - \bar{x},$ $\tilde{u} = u - \bar{u},$ $\Gamma(x,u) \triangleq [f(x_1, x_2, u_1)^\mathrm{T} \; g(x_2, u_2)^\mathrm{T}]^\mathrm{T},$ $A \triangleq \left.\frac{\partial\Gamma(x,u)}{\partial x}\right\rvert_{(\bar{x},\bar{u})}$ and $B \triangleq \left.\frac{\partial \Gamma(x,u)}{\partial u}\right\rvert_{(\bar{x},\bar{u})},$ 
the controllability matrix of the linearized system
\begin{align}
    \dot{\tilde{x}} = A\tilde{x} + B\tilde{u}, \quad
    0 =\Gamma(\bar{x},\bar{u}) \label{eq:linearized}
\end{align}
is not full rank, that is,
$\mathrm{rank}\left(\mathcal{C}\right) \neq n_1 + n_2,$
where $\mathcal{C} \triangleq [B \; AB \; A^2B \; \ldots \; A^{n_1+n_2-1}B].$
\end{assum}
\begin{rem}
The uncontrollability of the linearized system originates from Assumptions \ref{ass:overactuated} and \ref{ass:kernel}.
Since the system is overactuated, we can lose the ability to generate a nonzero $\tilde{u}_1$ for $u_1 \neq 0$ if the triple $(x_1,x_2,u_1)$ belongs to the kernel space.
As a consequence, we lose controllability in the linearized system.
\end{rem}

Let $x_{1,\mathrm{d}} \in \mathbb{R}^{n_1}$ and $x_{2,\mathrm{d}} \in \mathbb{R}^{n_2}$ be the desired states of $\mathrm{S}_1$ and $\mathrm{S}_2,$ respectively.
The \emph{primary} control objective is to make $x_1$ converge to $x_{1,\mathrm{d}}.$
Note that, for all $t\ge0,$ if $x_2=x_{2,\mathrm{d}},$ the dynamics of $\mathrm{S}_2$ can be neglected and the dynamics of $\mathrm{S}_1$ only depend on $x_1,$ $u_1,$ and $x_{2,\mathrm{d}}.$
%
In fact, assuming $x_2 = x_{2,\mathrm{d}}$ at all times, the primary control objective can be achieved by suitably designing $u_1$ such that $\mathrm{S}_1$ generates the \emph{ideal effective control}
\begin{align}\label{eq:ideal_Psi}
    \tilde{u}_{1,\mathrm{d}} \triangleq \Psi(x_1, x_{2,\mathrm{d}}, u_1),
\end{align}
where $\lim_{t\to\infty} x_1 = x_{1,\mathrm{d}}.$

%
However, the dynamics of $\mathrm{S}_2$ are not instantaneous. That is, the assumption $x_2 = x_{2,\mathrm{d}}$ does not hold at all times.
Interestingly, if $x_2$ converges ``faster" to $x_{2,\mathrm{d}}$ than $x_1$ to $x_{1,\mathrm{d}},$ then we can prove that the system is asymptotically stable. 
In particular, for $x_2\neq x_{2,\mathrm{d}},$ a disturbance is generated by $\mathrm{S}_2$ and enters $\mathrm{S}_1,$ and, reciprocally, a disturbance is generated by $\mathrm{S}_1$ and enters $\mathrm{S}_2.$
If the disturbances eventually vanish, it is possible to formally prove the system stability by using the small gain theorem \cite{Khalil:1173048}.

Additionally, note that this control problem is a two-level control-allocation problem \cite{JOHANSEN20131087}.
The secondary objective is dual and can be formulated as
(i) to design $u_2$ such that $x_2$ asymptotically converges to $x_{2,\mathrm{d}},$ and (ii) to design $x_2$ through an \emph{allocated mapping} to generate the ideal effective control\footnote{The ``ideal effective control" in this paper is called the ``\emph{commanded virtual control}" in \cite{JOHANSEN20131087}.}. 
%

\begin{figure}
    \centering
    \includegraphics[width=0.4\columnwidth]{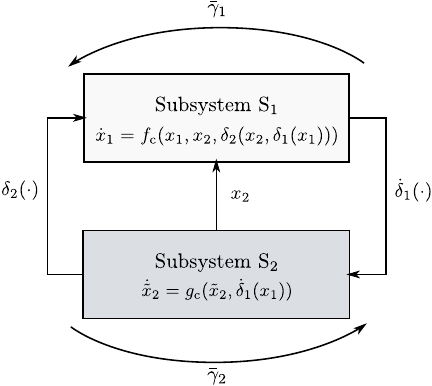}
    \caption{
    Cascaded closed-loop system \eqref{eq:S1_controlled}, \eqref{eq:S2_controlled}.
    %
    }
    \label{fig:closed_loop_sys}
\end{figure}

Now, assume that a continuous control-allocation law for \eqref{eq:S1}, \eqref{eq:S2} is designed such that the closed-loop system has the cascaded form
\begin{align}
    \dot{x}_1 &= f_\mathrm{c}(x_1, x_2, \delta_2(x_2, \delta_1(x_1))), \label{eq:S1_controlled} \\
    \dot{\tilde{x}}_2 &= g_\mathrm{c}(\tilde{x}_2, \dot{\delta}_1(x_1)), \label{eq:S2_controlled}
\end{align}
where $\tilde{x}_2\in\mathbb{R}^{n_2}$ is the error\footnote{
The definition of $\tilde{x}_2$ depends on the metric used by the space of $x_2.$ For example, in the Euclidian space, a judicious choice of the error is $\tilde{x}_2 \triangleq x_2 - x_{2,\mathrm{d}},$ whereas on $SO(3),$ that is, for $x_2\in SO(3),$ $\tilde{x}_2 \triangleq x_2^\mathrm{T} x_{2,\mathrm{d}}$ is preferable.}
between $x_2$ and $x_{2,\mathrm{d}},$ $\delta_1:\mathbb{R}^{n_1}\to\mathbb{R}^{n_2}: x_{2,\mathrm{d}} = \delta_1(x_1)$ is the allocated mapping from $x_1$ to $x_{2,\mathrm{d}},$ $\dot{\delta}_1(\cdot)$ is the exogenous input of $\mathrm{S}_2$ and is the time derivative of the allocated mapping, $\delta_2:\mathbb{R}^{n_1}\times\mathbb{R}^{n_2} \to \mathbb{R}^{n_2}: \tilde{x}_2 = \delta_2(x_2, \delta_1(x_1))$ is the exogenous input of $\mathrm{S}_1$ and is the state error produced by the dynamics of $\mathrm{S}_2,$ and $f_\mathrm{c}:\mathbb{R}^{n_1}\times\mathbb{R}^{n_2}\times\mathbb{R}^{n_2}\to\mathbb{R}^{n_1}$ and $g_\mathrm{c}:\mathbb{R}^{n_2}\times\mathbb{R}^{n_2}\to\mathbb{R}^{n_2}$ are continuously differentiable, globally Lipschitz, and nonlinear.
The closed-loop system is depicted in Figure \ref{fig:closed_loop_sys}.
Note that this control law introduces the allocated mapping $\delta_1(\cdot),$ and the two exogenous inputs $\dot{\delta}_1(\cdot)$ and $\delta_2(\cdot).$
The canonical formulations \eqref{eq:S1}, \eqref{eq:S2} and \eqref{eq:S1_controlled}, \eqref{eq:S2_controlled} extend the results of \cite{NGUYEN2021109586}.
At first sight, these canonical forms seem impractical, but
we will show in the next section that a relevant class of thrust vectoring systems can be formulated as is. 
Other relevant examples can be found in
\cite{doi:10.2514/1.G004356},
\cite{NGUYEN2019384},
\cite{NGUYEN2021109586},
and \cite{NICOTRA2017174}.
Next, assume the following.

\begin{assum}\label{ass:exp_stability}
%
Let $\tilde{x}_1 \triangleq x_1 - x_{1,\mathrm{d}}$ and 
consider the closed-loop system \eqref{eq:S1_controlled}, \eqref{eq:S2_controlled} rewritten as 
\begin{align}
    \dot{\tilde{x}}_1 &= \tilde{f}_\mathrm{c}(\tilde{x}_1, \tilde{\delta}_2(\tilde{x}_2)), \label{eq:S1_controlled_shifted} \\
    \dot{\tilde{x}}_2 &= \tilde{g}_\mathrm{c}(\tilde{x}_2, \dot{\tilde{\delta}}_1(\tilde{x}_1)). \label{eq:S2_controlled_shifted}
\end{align}
%
%
In the absence of exogenous inputs, that is, for $\dot{\tilde{\delta}}_1(\cdot) = \tilde{\delta}_2(\cdot) = 0,$ the allocated mapping $\tilde{\delta}_1(\cdot)$ and the ideal effective control law \eqref{eq:ideal_Psi} make the origin of \eqref{eq:S1_controlled_shifted} and \eqref{eq:S2_controlled_shifted} exponentially stable, where $\lim_{t\to\infty} \tilde{x}_i = 0 \; (i\in\{1,2\}).$
That is, for $i\in\{1,2\},$ let $\tilde{x}_{i} = 0$ be an equilibrium point and $D_{S_i} \subset \mathrm{R}^{n_i}$ be a domain containing the equilibrium. Then, for $i\in\{1,2\},$ we assume that $S_i$ has a Lypunov function candidate $V_{S_i} : [0,\infty) \times D_{S_i} \to \mathbb{R}$ such that, $\forall t\in\mathbb{R}_{\geq0}$ and $\forall \tilde{x}_i \in D_{S_i},$
\begin{align}
\underline{W}_{S_i}(\tilde{x}_i) \leq V_{S_i}(t, \tilde{x}_i) \leq \bar{W}_{S_i}(\tilde{x}_i), \\
\dot{V}_{S_i}(t, \tilde{x}_i) \leq - \bar{\bar{W}}_{S_i}(\tilde{x}_i),
\end{align}
where $\underline{W}_{S_i}(\tilde{x}_i) \geq \kappa_{1,i} \|\tilde{x}_i\|^{\zeta_i},$ $\bar{W}_{S_i}(\tilde{x}_i) \leq \kappa_{2,i} \|\tilde{x}_i\|^{\zeta_i},$ and $\bar{\bar{W}}_{S_i}(\tilde{x}_i) \geq \kappa_{3,i} \|\tilde{x}_i\|^{\zeta_i}$ are positive definite functions on $D_{S_i}$ and $\kappa_{1,i}, \kappa_{2,i}, \kappa_{3,i}, {\zeta_i} \in\mathbb{R}_{>0}$ are positive constants.
\end{assum}

\begin{assum}\label{ass:iss}
In the presence of the bounded exogenous inputs $\dot{\tilde{\delta}}_1(\cdot)$ and $\tilde{\delta}_2(\cdot),$ $\mathrm{S}_1$ and $\mathrm{S}_2$ are input-to-state stable (ISS) with respect to $\dot{\tilde{\delta}}_1(\cdot)$ and $\tilde{\delta}_2(\cdot),$ respectively.
That is, there exist a class $\kappa\ell$ function $\beta_1,$ a class $\kappa\ell$ function $\beta_2,$ a class $\kappa$ function $\gamma_1,$ a class $\kappa$ function $\gamma_2,$ and positive constants $k_1, k_2, \kappa_1, \kappa_2 \in \mathbb{R}_{\ge0}$ such that, for all $\tilde{x}_1(t_0)$ and $\tilde{x}_1(t_0)$ with $\|\tilde{x}_1(t_0)\| < k_1$ and $\|\tilde{x}_2(t_0)\| < \kappa_1,$ and for all $\dot{\tilde{\delta}}_1$ and $\tilde{\delta}_2$ with $\sup_{t\geq t_0} \|\dot{\tilde{\delta}}_1\| < k_2$ and $\sup_{t\geq t_0} \|\dot{\tilde{\delta}}_2\| < \kappa_2,$ the solutions $\tilde{x}_1$ and $\tilde{x}_2$ satisfy
\begin{align}
    \|\tilde{x}_i\| & \leq \beta_i(\|\tilde{x}_i(t_{0,i})\|, \tilde{t}_i) + \gamma_i\left(\sup_{\tau\in[t_{0,i},t]}\|\tilde{\delta}_j(\tau)\|\right), \label{eq:xi_iss}
\end{align}
for all $t\ge t_{0,i},$ where $i\in\{1,2\},$ $j\neq i \in\{1,2\},$ and $\tilde{t}_i = t-t_{0,i}.$
%
\end{assum}

%

%
\begin{rem}
According to \eqref{eq:xi_iss}, since the exogenous inputs are bounded, for $t \to \infty,$ it follows from ISS that the states are eventually bounded by
\begin{align}
\lim_{t\to\infty} \|\tilde{x}_i\| \leq \gamma_i\left(\sup_{\tau\in[0,\infty)} \|\tilde{\delta}_j(\tau)\|\right),
\end{align}
where $i\in\{1,2\}$ and $j\neq i \in \{1,2\}.$
As a result, since $\tilde{\dot{\delta}}_1(\cdot)$ and $\tilde{\delta}_2(\cdot)$ are bounded and are functions of $\tilde{x}_1$ and $\tilde{x}_2,$ respectively, it follows that there exists $T \in \mathbb{R}_{>0}$ such that, for all $t\ge T,$
\begin{align}
    \|\dot{\tilde{\delta}}_1\| \leq \bar{\gamma}_1 \|\tilde{\delta}_2\|, \quad
\|\tilde{\delta}_2\| \leq \bar{\gamma}_2 \|\dot{\tilde{\delta}}_1\|, \label{eq:iss_delta_as}
\end{align}
where $\bar{\gamma}_1\in\mathbb{R}_{>0}$ is the asymptotic gain between the output $\dot{\tilde{\delta}}_1(\cdot)$ and the input $\tilde{\delta}_2(\cdot),$ and $\bar{\gamma}_2\in\mathbb{R}_{>0}$ is the asymptotic gain between the output $\tilde{\delta}_2(\cdot)$ and the input $\dot{\tilde{\delta}}_1(\cdot).$
\end{rem}

Fig. \ref{fig:closed_loop_sys} depicts the complete structure of the interconnected system with the asymptotic gains.
%
Under Assumptions \ref{ass:exp_stability} and \ref{ass:iss}, it is possible to prove the uniform asymptotic convergence of the state to the desired one using small gain arguments.

\begin{thm}\label{th:UAS}
Consider the closed-loop system \eqref{eq:S1_controlled_shifted}, \eqref{eq:S2_controlled_shifted}.
If the control law satisfies Assumption \ref{ass:exp_stability}, and the mapping $\delta_1(\cdot)$ is Lipschitz continuous, satisfies the input constraint, and is designed such that Assumption \ref{ass:iss} holds, then the closed-loop system is uniformly asymptotically stable (UAS) for $\bar{\gamma}_2$ suitably small such that $\bar{\gamma}_1 \bar{\gamma}_2 < 1.$
Furthermore, the states asymptotically converge to their desired state, that is, $\lim_{t\to\infty}\tilde{x}_1 = 0$ and $\lim_{t\to\infty}\tilde{x}_2 = 0.$
\end{thm}
\begin{pf}
First, note that, for any initial condition that is not an equilibrium, the allocated mapping $\tilde{\delta}_1(\cdot)$ evolves with the trajectories of \eqref{eq:S1_controlled_shifted} to achieve the primary control objective. Thus, during the transients, $\tilde{\delta}_1(\cdot)$ is time-varying, that is, $\|\dot{\tilde{\delta}}_1\| \ge0$.
Next, since the dynamics of \eqref{eq:S2_controlled_shifted} are not instantaneous and $\tilde{\delta}_2 \triangleq \tilde{x}_2,$  note that $\|\tilde{\delta}_2\| \ge 0$ is generated by \eqref{eq:S2_controlled_shifted} during the transients.
As a result, the interconnected system generates two exogenous inputs entering each subsystem for any initial condition that is not an equilibrium.

According to Assumption \ref{ass:iss}, \eqref{eq:S1_controlled_shifted}, \eqref{eq:S2_controlled_shifted} are ISS with respect to $\tilde{\delta}_2$ and $\dot{\tilde{\delta}}_1,$ respectively.
That is, according to \eqref{eq:xi_iss}, the trajectories of $\tilde{x}_1$ and $\tilde{x}_2$ are contained within a ball whose radius is a $\kappa$ function of the supremum of $\|\tilde{\delta}_2\|$ and $\|\dot{\tilde{\delta}}_1\|,$ respectively.
Clearly, this property holds true only if the exogenous inputs are bounded.
To enforce this condition, as the control law and the system dynamics are continuous, the allocated mapping $\tilde{\delta}_1$ must be Lipschitz continuous to bound its time derivative.

Next, we prove that, if the small gain \cite{Khalil:1173048} condition $\bar{\gamma}_1 \bar{\gamma}_2 < 1$ holds true, the exogenous inputs eventually vanish.
This is done by analyzing the asymptotic behavior of the exogenous inputs in \eqref{eq:iss_delta_as} and comparing the input signal with the output signal after one loop iteration.
In particular, note that $\tilde{\delta}_2$ ``loops" through the interconnected system by traversing $S_1$ and $S_2$ (see Fig. \ref{fig:closed_loop_sys}). The following argument compares the input signal $\tilde{\delta}_2$ with the output signal $\tilde{\delta}'_2$ after one loop iteration.
In particular, using \eqref{eq:iss_delta_as}, we have
\begin{align}
    \|\tilde{\delta}'_2\| \le \bar{\gamma}_1 \bar{\gamma}_2 \|\tilde{\delta}_2\|,
\end{align}
and, since $\bar{\gamma}_1\bar{\gamma}_2 < 1,$ it is necessary that
$\|\tilde{\delta}'_2\| < \|\tilde{\delta}_2\|,$
which proves that the norm of $\tilde{\delta}_2$ asymptotically decreases.
The same arguments hold true for $\dot{\tilde{\delta}}_1.$

Therefore, since the exogenous inputs eventually vanish and, according to Assumption \ref{ass:exp_stability}, both subsystems are exponentially stable in the absence of exogenous inputs, UAS\footnote{Note that, as a direct consequence of the exponential stability of Assumption \ref{ass:exp_stability}, $\beta_1(\cdot)$ and $\beta_2(\cdot)$ in \eqref{eq:xi_iss}, respectively, have an exponential form. However, because the exogenous inputs do not decrease exponentially, it is only possible to claim UAS (not exponential stability) of the origin of the interconnected system.} follows and the state error converges to zero, which completes the proof.
%

\end{pf}



\section{Case Study: Planar Control of a UAV Manipulating an Object}\label{sec:case_study}
\begin{figure}
    \centering
    \includegraphics[width=0.3\columnwidth]{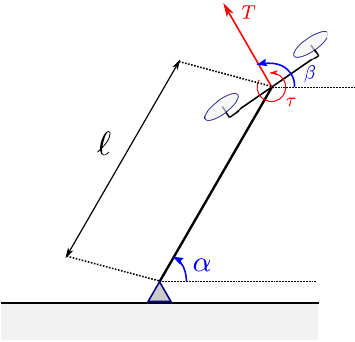}
    \caption{UAV manipulating an object in two dimensions.
    }
    \label{fig:UAV_manipulating_object}
\end{figure}

Consider the planar model of a UAV manipulating an object as depicted in Figure \ref{fig:UAV_manipulating_object}.
Assume that the center of mass of the UAV concides with the joint position, and the center of mass of the object is at half distance between the UAV and the ground joint.
Let $m_\mathrm{u}\in\mathbb{R}_{>0}$ be the UAV mass,
$I_\mathrm{u}\in\mathbb{R}_{>0}$ the UAV inertia,
$m_{\mathrm{o}}\in\mathbb{R}_{>0}$ the object mass,
$I_\mathrm{o}\in\mathbb{R}_{>0}$ the object inertia,
and $\ell\in\mathbb{R}_{>0}$ the object length.
Additionally, let $\alpha\in[0,\pi]$ be the angle of the object with respect to the horizontal,
and $\beta\in[-\pi, \pi)$ the angle between the UAV thrust and the horizontal.
The system is actuated by the UAV torque $\tau\in\mathbb{R}$ and the UAV thrust $T\in\mathbb{R}_{\ge0}.$

Using the Euler-Lagrange method, the equations of motion are given by \cite{NGUYEN2019384}
\begin{align}
    \tilde{I} \ddot\alpha + \tilde{m}\ell g\cos\alpha & = T\ell\sin(\beta-\alpha), \label{eq:UAV_sys1} \\
    I_\mathrm{u}\ddot\beta &= \tau, \label{eq:UAV_sys2}
\end{align}
where $\tilde{I} \triangleq \frac{m_\mathrm{o} \ell^2}{4} + I_\mathrm{o} + m_\mathrm{u}\ell^2,$ $\tilde{m} \triangleq \frac{m_\mathrm{o}}{2} + m_\mathrm{u},$ and $g\in\mathbb{R}_{>0}$ is the gravity acceleration.

Let $x_1 \triangleq [x_{1,1} \; x_{1,2}]^\mathrm{T} = [\alpha \; \dot{\alpha}]^\mathrm{T},$
$x_2 = [x_{2,1} \; x_{2,2}]^\mathrm{T} = [\beta \; \dot{\beta}]^\mathrm{T},$
$u_1 = T,$ and $u_2 = \tau.$
The system \eqref{eq:UAV_sys1}, \eqref{eq:UAV_sys2} can be rewritten in the canonical form \eqref{eq:S1}, \eqref{eq:S2} as
\begin{align}
    \dot{x}_1 &= \left[
    \begin{smallmatrix}
    x_{1,2} \\
    \tilde{I}^{-1}\ell(u_1\sin(x_{2,1} - x_{1,1}) - \tilde{m} g \cos x_{1,1})
    \end{smallmatrix}
    \right] \nonumber\\
    &= f(x_1, x_2, u_1), \label{eq:UAV_sys_canonical1} \\
    \dot{x}_2 &= \left[
    \begin{smallmatrix}
    x_{2,2} \\
    I_\mathrm{u}^{-1}u_2
    \end{smallmatrix}
    \right] = g(x_2, u_2). \label{eq:UAV_sys_canonical2}
\end{align}

Using \eqref{eq:Psi}, the effective control can be formulated as
\begin{align}
    \tilde{u}_1 & \triangleq \Psi(x_1, x_2, u_1) = u_1 \sin(x_{2,1} - x_{1,1}).
\end{align}
Note that the system is overactuated since $\dim(u_1) + \dim(x_2) = 3 > \dim(\tilde{u}_1) = 1,$ which complies with Assumption \ref{ass:overactuated}.
Additionally, note that, according to \eqref{eq:kernel}, $\ker(\Psi(\cdot))$ is given by
\begin{align}\label{eq:kernelUAV}
    \ker(\Psi(\cdot)) \hspace{-0.25em}= \hspace{-0.25em} \{(x_1,x_2,u_1):u_1\sin(x_{2,1}-x_{1,1}) = 0\}.
\end{align}
From \eqref{eq:kernelUAV}, we can see that, for all $x_1\in\mathbb{R}^{n_1}$ and for all $u_1^*\in\mathbb{R}^{m_1},$ there exists a class $C^0$ function $\mathcal{K}(\cdot)$ such that $x^*_2 = \mathcal{K}(x_1) \Rightarrow \Psi(x_1,x_2^*,u_1^*) = 0.$
In particular, $\mathcal{K}(\cdot)\in\mathcal{S}_\mathcal{K},$ where, for all $\gamma\in\mathbb{R}$ and for all $n\in\mathbb{Z},$
$
\mathcal{S}_\mathcal{K} \triangleq \left\{\mathcal{K}_n:\mathbb{R}^2\to\mathbb{R}^2: \mathcal{K}_n(\cdot) = \left[
x_{1,1} + n\pi \;
\gamma \right]^\mathrm{T}\right\}.
$
Note that this fact complies with Assumption \ref{ass:kernel}.

Next, consider the equilibrium point 
$\bar{x}_1 = \bar{x}_2 = [\frac{\pi}{2} \; 0]^\mathrm{T}$
and $\bar{u}_1 = \bar{u}_2 = 0.$
%
%
The system linearized around $(\bar{x}, \bar{u}) \triangleq ([\bar{x}_1^\mathrm{T} \; \bar{x}_2^\mathrm{T}]^\mathrm{T}, [\bar{u}_1 \; \bar{u}_2]^\mathrm{T})$ is given by \eqref{eq:linearized}, where
$
    A = \left[
    \begin{smallmatrix}
    0 & 1 & 0 & 0 \\
    \tilde{I}^{-1}\ell\tilde{m}g & 0 & 0 & 0 \\
    0 & 0 & 0 & 1 \\
    0 & 0 & 0 & 0
    \end{smallmatrix}\right], 
    B = \left[
    \begin{smallmatrix}
    0 & 0 \\
    0 & 0 \\
    0 & 0 \\
    0 & I_\mathrm{u}^{-1}
    \end{smallmatrix}
    \right].
$
The rank of the controllability matrix is computed by
$
    \mathrm{rank}(\mathcal{C}) = \mathrm{rank}\left(\left[
    \begin{smallmatrix}
    0 & 0 & 0 & 0 & 0 & 0 & 0 & 0 \\
    0 & 0 & 0 & 0 & 0 & 0 & 0 & 0 \\
    0 & 0 & 0 & I_\mathrm{u}^{-1} & 0 & 0 & 0 & 0 \\
    0 & I_\mathrm{u}^{-1} & 0 & 0 & 0 & 0 & 0 & 0 
    \end{smallmatrix}
    \right]\right) = 2 < 4,
$
which is not full, and thus implies that the linearized system is uncontrollable.
This fact complies with Assumption \ref{ass:controllability}.
Note that this important equilibrium point is meaningful, as it represents the UAV at its minimum energy (zero torque and zero thrust), when the object is positioned vertically.
In this case, there is not much we can do to stabilize the system with a linear control law.
Hence, we need to come up with a nonlinear control law and work directly with \eqref{eq:UAV_sys_canonical1}, \eqref{eq:UAV_sys_canonical2}.

The primary control objective is to stabilize the object angle to the desired angle $\alpha_{\mathrm{d}} \in [0, \pi].$
Note that the desired UAV attitude $\beta_{\mathrm{d}} \in [-\pi,\pi)$ needs to be suitably designed to achieve the primary control objective.
In \cite{NGUYEN2019384}, we propose the control law
\begin{align}
    u_1 &= \frac{\tilde{u}_{1,\mathrm{d}}}{\sin(\beta_\mathrm{d} - \alpha)}, \;\;
    u_2 = k_{\mathrm{p},\beta}(\beta_\mathrm{d}-\beta) - k_{\mathrm{d},\beta} \dot{\beta}, \label{eq:control_law1} \\
    \tilde{u}_{1,\mathrm{d}} &= k_{\mathrm{p},\alpha}(\alpha_\mathrm{d}-\alpha) - k_{\mathrm{d},\alpha}\dot\alpha + \tilde{m} g \cos\alpha, \label{eq:control_law3} \\
    \beta_\mathrm{d} &\triangleq \arctan(\varepsilon \tilde{u}_{1,\mathrm{d}}) + \alpha, \label{eq:control_law4}
\end{align}
where $k_{\mathrm{p},\alpha} > 0,$ $k_{\mathrm{d},\alpha} > 0,$ $k_{\mathrm{p},\beta} > 0,$ $k_{\mathrm{d},\beta} > 0,$ and $\varepsilon>0$ are the control parameters.
Note that, according to the definition \eqref{eq:ideal_Psi}, \eqref{eq:control_law3} is the ideal effective control.
Furthermore, \eqref{eq:control_law4} is the proposed allocated mapping, which is Lipschitz continuous in $x_1$ and ensures that $u_1\ge0.$
Without further details, note that \eqref{eq:control_law4} has been derived by taking advantage of $\ker(\Psi(\cdot))$ to smooth out the allocated mapping.

Let 
$x_{1,\mathrm{d}} \triangleq [x_{1,1,\mathrm{d}} \; 0]^\mathrm{T}$
and
$x_{2,\mathrm{d}} \triangleq [x_{2,1,\mathrm{d}} \; 0]^\mathrm{T}$
be the desired states,
$\tilde{x}_2\triangleq x_{2,\mathrm{d}} - x_2 = [(x_{2,1,\mathrm{d}} - x_{2,1}) \; x_{2,2}]^\mathrm{T}$ be the attitude and the angular-velocity error, 
and define
$
    \delta_{1}(x_1) \triangleq x_{2,\mathrm{d}} = [\arctan(\varepsilon \tilde{u}_{1,\mathrm{d}}) \; 0]^\mathrm{T},
    %
    \delta_2(x_2, \delta_1(x_1)) \triangleq \tilde{x}_2 = \delta_1(x_1) - x_{2}.
$
Using $\tilde{\delta}_1 \triangleq [1 \; 0]\delta_1(\cdot),$ and $\tilde{\delta}_2 \triangleq [1 \; 0]\delta_2(\cdot),$ we can transform \eqref{eq:UAV_sys_canonical1}, \eqref{eq:UAV_sys_canonical2} with \eqref{eq:control_law1}-\eqref{eq:control_law4} to \eqref{eq:S1_controlled}, \eqref{eq:S2_controlled} as
\begin{align}
    \dot{x}_1 &= \left[
    \begin{smallmatrix}
    x_{1,2} \\
    \frac{\ell}{\tilde{I}}(\frac{\tilde{u}_{1,\mathrm{d}}}{\sin(x_\mathrm{2,1,d}-x_{1,1})}\sin(x_{2,1,\mathrm{d}} - \tilde\delta_2 - x_{1,1}) - \tilde{m} g \cos x_{1,1})
    \end{smallmatrix}
    \right] \nonumber\\
    &= f_\mathrm{c}(x_1, x_2, \delta_2(x_2,\delta_1(x_1))), \label{eq:UAV_sys_canonical3} \\
    \dot{\tilde{x}}_2 &= \left[
    \begin{smallmatrix}
    \dot{\tilde\delta}_1 - \tilde{x}_{2,2} \\
    I_\mathrm{u}^{-1}(k_{\mathrm{p},\beta}\tilde{x}_{2,1} - k_{\mathrm{d},\beta} \tilde{x}_{2,2})
    \end{smallmatrix}
    \right] = g_\mathrm{c}(\tilde{x}_{2}, \dot{\delta}_1(x_1)). \label{eq:UAV_sys_canonical4}
\end{align}

According to Lemma 2 and Lemma 4 of \cite{NGUYEN2019384}, the closed-loop subsystems \eqref{eq:UAV_sys_canonical3} and \eqref{eq:UAV_sys_canonical4} are ISS with respect to $\delta_2(\cdot)$ and $\dot{\delta}_1(\cdot),$ respectively.
Note that $\bar{\gamma}_2$ can be made arbitrarily small for suitably large gains $k_{\mathrm{p},\beta}$ and $k_{\mathrm{d},\beta}.$
Hence, since Assumptions \ref{ass:exp_stability} and \ref{ass:iss} hold true, Theorem \ref{th:UAS} follows.
Thus, the closed-loop system \eqref{eq:UAV_sys_canonical3}, \eqref{eq:UAV_sys_canonical4} is UAS.

\begin{figure}
    \centering
    \includegraphics[width=.5\columnwidth, trim={0.7cm 1.2cm 0.7cm 0.7cm}]{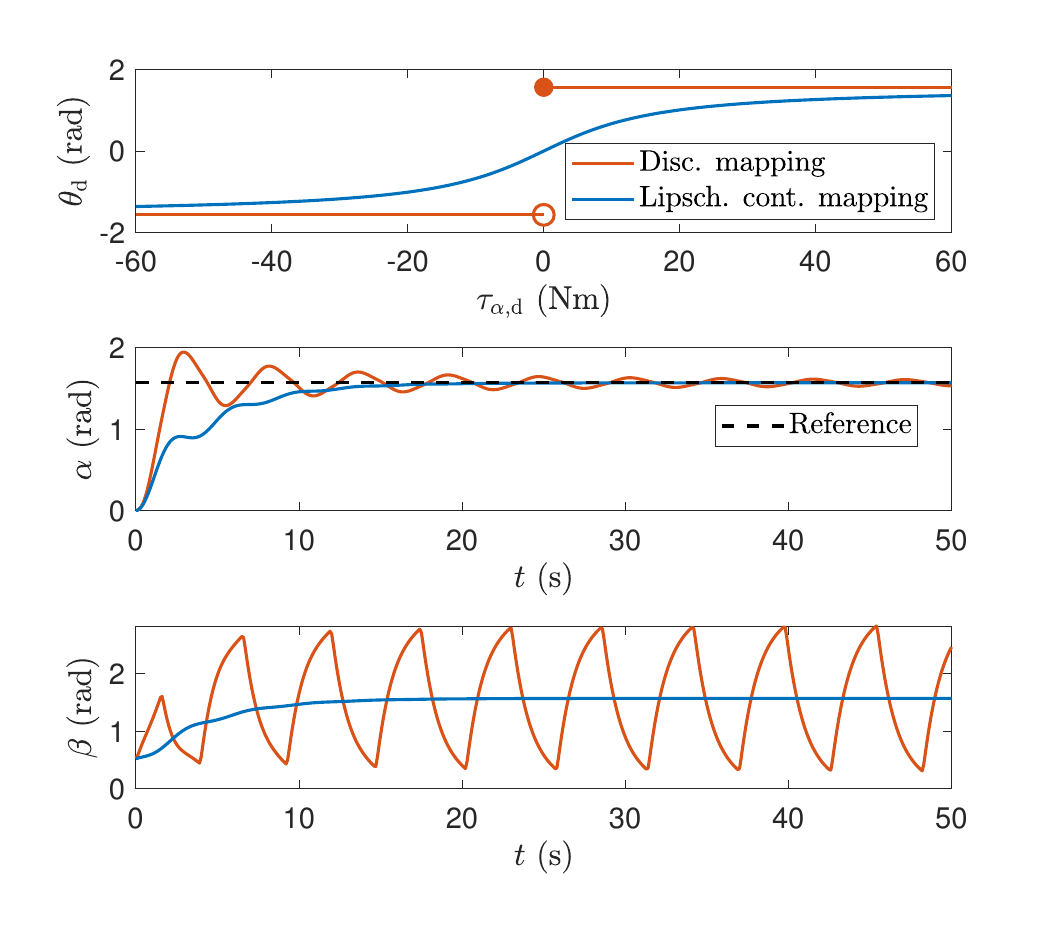}
    \caption{Comparison between the optimal mapping \eqref{eq:optimal_mapping} and the Lipschitz continuous mapping \eqref{eq:control_law4}. We see that the optimal but discontinuous mapping cannot stabilize the system.}
    \label{fig:uav_case_study}
\end{figure}

The optimal mapping can be obtained by following the same steps as in \cite{1386824}, which uses the pseudo inverse in the general case.
According to \eqref{eq:UAV_sys1}, the desired torque applied to the ground joint is computed by $\tau_{\alpha, \mathrm{d}} \triangleq \ell\tilde{u}_{1,\mathrm{d}} = \tau_\alpha\sin(\theta),$ where $\tau_\alpha \triangleq T\ell$ and $\theta \triangleq \beta - \alpha.$
To obtain the optimal mapping, we must ensure that the desired and applied torques match, that is, $\tau_{\alpha,\mathrm{d}} = \tau_\alpha,$ so that there are no losses. Therefore, since $T\ge0$ and $\ell>0,$ we must enforce that $\sin(\theta) = 1$ for $\tau_{\alpha,\mathrm{d}} > 0$ and $\sin(\theta) = -1$ for $\tau_{\alpha,\mathrm{d}} < 0,$ where the mapping is undefined for $\tau_{\alpha,\mathrm{d}} = 0.$
Accordingly, the optimal mapping is chosen as
\begin{align}
    \theta_\mathrm{d} &= \begin{cases}
    \pi/2, & \mathrm{if} \; \tau_{\alpha,\mathrm{d}} \ge 0, \\
    -\pi/2, & \mathrm{if} \; \tau_{\alpha,\mathrm{d}} < 0,
    \end{cases}
    \label{eq:optimal_mapping}
\end{align}
where $\theta_\mathrm{d} \triangleq \beta_\mathrm{d} - \alpha.$
Clearly, \eqref{eq:optimal_mapping} is discontinuous at $\tau_{\alpha,\mathrm{d}}=0$ compared to \eqref{eq:control_law4}, which can be rewritten as $\theta_\mathrm{d} = \arctan(\varepsilon\ell^{-1} \tau_{\alpha,\mathrm{d}}).$
Using the same parameters of Example \ref{subsec:example1}, Figure \ref{fig:uav_case_study} shows that, since \eqref{eq:optimal_mapping} is not Lipschitz continuous, the system exhibits oscillatory behaviors.

%
In this section, it was seen that, since the system could lose controllability when linearized around a particular point of equilibrium, it was not trivial to design a nonlinear control law.
Furthermore, we had to ensure that the UAV thrust was positive at all times.
It was shown that the generalized-inverse method led to a discontinuous mapping, which made the system oscillate.
Accordingly, we proposed a nonlinear control law using the analytic Lipschitz-continuous mapping \eqref{eq:control_law4}.
However, as discussed in \cite{NGUYEN2021109586}, finding such a closed-form mapping is in general nontrivial and not unique.
In fact, in a lot of situations, finding such an analytic mapping is quite hard.
Therefore, in this paper, we propose an alternative numerical solution and assess if we can substitute the design of a closed-form mapping by solving an optimization problem online.
%

\section{Kernel-Based Predictive Control Allocation}

In this section, we propose the new kernel-based predictive control allocation, which takes advantage of the kernel of $\Psi(\cdot)$ in \eqref{eq:Psi} to numerically compute a locally smooth allocated mapping.
 Since KPCA substitutes the need for designing $\delta_1(\cdot)$ by solving an optimization problem at each time step, we alleviate the difficulty of finding a closed-form solution for the mapping.
%

%
%

For all $x\in\mathbb{R},$ define the Gaussian bell-curve function
\begin{align}\label{eq:bell_curve}
    \Omega(x, \kappa_\mathrm{p}, \kappa_\mathrm{w}) \triangleq \kappa_\mathrm{p}\exp{\left(-x^2/\kappa_\mathrm{w}\right)},
\end{align}
where $\kappa_\mathrm{p}\in\mathbb{R}_{>0}$ and $\kappa_\mathrm{w}\in\mathbb{R}_{>0}$ are the bell-curve parameters.
Note that $\Omega(\cdot)$ is symmetric, and $\kappa_\mathrm{p}$ and $\kappa_\mathrm{w}$ modify the bell-curve peak and width, respectively.
Additionally, for all $\kappa_\mathrm{p}>0$ and $\kappa_\mathrm{w}>0,$ $\Omega(\cdot)$ reaches its maximum at $\Omega(0, \kappa_\mathrm{p}, \kappa_\mathrm{w}) = \kappa_\mathrm{p}.$
%

%
Let $t_\mathrm{N} \in \mathbb{R}_{>0}$ be the time length of the prediction horizon. For all $t \in \mathbb{R}_{\ge 0},$ the continuous-time NLP problem is formulated as
\begin{align}
\min_{u, x_{2, \mathrm{d}}}J =& \int_{t}^{t + t_{\mathrm{N}}} [(x-x_\mathrm{d})^\mathrm{T}Q(x-x_\mathrm{\mathrm{d}})
+ \dot{u}^\mathrm{T}R \dot{u} \nonumber \\
& \;\;\;\;\;\;\;\;  + \Omega(\|\tilde{u}_{1,\mathrm{d}}\|_2, \kappa_\mathrm{p}, \kappa_\mathrm{w}) \|\tilde{\mathcal{K}}\|^2_2] dt , \nonumber \\
\mathrm{s.t.} & \quad \quad \dot{x}_1 = f(x_1, x_2, u_1), \; \dot{x}_2 = g(x_2, u_2), \nonumber
\\
& \quad \quad u_{\mathrm{min}} \le u \le u_{\mathrm{max}}, \; \Pi(x_{2,\mathrm{d}}) = 0, \label{eq:NLP}
\end{align}
where
$x_{\mathrm{d}} = [x^\mathrm{T}_{1, \mathrm{d}} \; x^{\mathrm{T}}_{2, \mathrm{d}}]^\mathrm{T}$ is the desired state,
$\tilde{\mathcal{K}}\triangleq x_{2,\mathrm{d}} - \mathcal{K}(x_1)$ is the error between $x_{2,\mathrm{d}}$ and $\mathcal{K}(x_1),$
%
%
$\tilde{u}_{1,\mathrm{d}}$ is the ideal effective control,
$\|\cdot\|_2$ is the $\ell_2$ norm,
$Q\in\mathbb{R}^{(n_1+n_2)\times(n_1+n_2)}_{>0}$ and $R\in\mathbb{R}^{(m_1+m_2)\times(m_1+m_2)}_{\ge0}$ 
are the state and the differential control weights, respectively,
and $\Pi(x_{2,\mathrm{d}}):\mathbb{R}^{n_2}\to\mathbb{R}$ is the equality constraint on $x_{2,\mathrm{d}}.$
Note that the inequality constraint is applied element-wise to $u.$
Furthermore, note that the desired state $x_{2, \mathrm{d}}$ is a decision variable, which substitutes the need for designing the allocated mapping $x_{2,\mathrm{d}} = \delta_1(\cdot).$
%
%
%
%
The key component of \eqref{eq:NLP} is the third summation in the cost function $J.$ Note that this summation vanishes as $\kappa_\mathrm{p}\to0.$
The goal is to minimize across the prediction horizon the $\ell_2$ norm of the error between $x_{2,\mathrm{d}}$ and a point in the kernel space of $\Psi(\cdot)$ as $\|\tilde{u}_{1,\mathrm{d}}\|_2$ approaches zero.
By doing so, $x_{2,\mathrm{d}}$ navigates close to a set of ``smoothing" points belonging to the kernel space when no effective control is needed.
%
%
Note that the deviation penalty from the kernel space can be tuned through $\kappa_\mathrm{p}$ and $\kappa_\mathrm{w}.$
Next, by solving \eqref{eq:NLP}, KPCA approximates ``on-the-fly" an allocated mapping, which is locally smooth in the vicinity of the kernel space to control the system.
%


Note that, as this paper does not provide the proofs of the closed-loop stability of KPCA, we demonstrate its stability through the next three numerical examples, which comply with the assumptions of the paper.
However, it is worth noting that the results of the previous sections are crucial, as they are the starting point and the intuition behind the design of KPCA. In fact, \eqref{eq:NLP} tries to ``mimic" the Lipschitz continuity of the analytic allocated mapping by using the kernel space, and the cascaded scheme is mimicked by the introduction of the allocated mapping as a decision variable.

\section{Simulations}\label{sec:sim}

In this section, we use the open-source software CasADI \cite{Andersson2019} to solve the NLP.
Denote by $k \in \mathbb{N}_0$ the $k$-th step of the discretized system, which corresponds to the time $t = kT_{\mathrm{s}},$ where $T_{\mathrm{s}} \in \mathbb{R}_{>0}$ is the sampling time. The prediction horizon in discrete time is denoted by $N\in\mathbb{N}.$ To numerically solve \eqref{eq:NLP} at each time step, the continuous-time system \eqref{eq:S1}, \eqref{eq:S2} is discretized using the fourth-order Runge-Kutta method \cite{suli_mayers_2003} and
the interior-point optimizer (IPOPT) \cite{Wachter2006} is used with warm start.

\subsection{Example 1: UAV manipulating an object in 2D}\label{subsec:example1}
\begin{figure}
    \centering
    \includegraphics[width=.45\columnwidth, trim={1cm 1.8cm 1cm 1.3cm}]{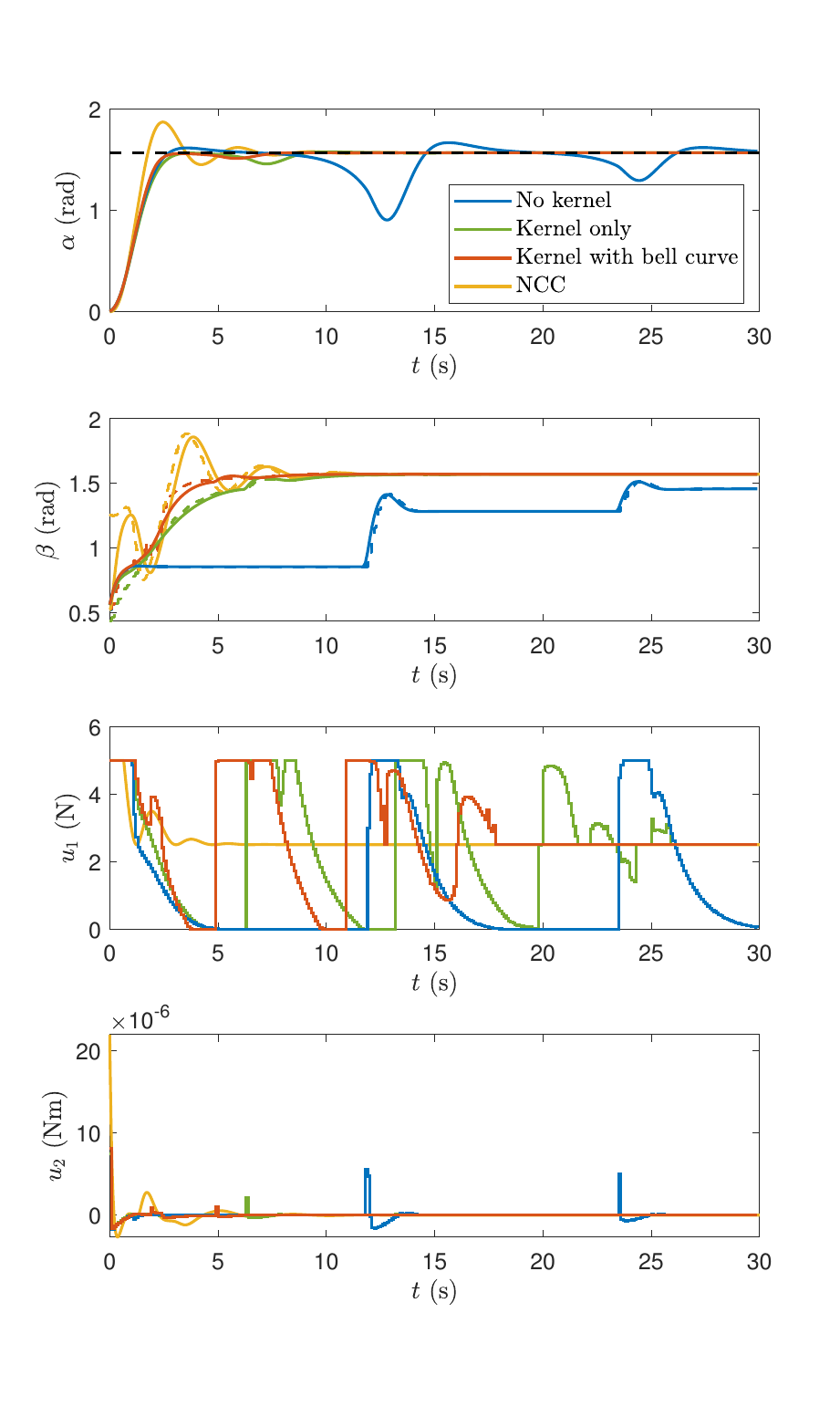}
    \caption{\textbf{Example 1.}
    One-step command following using $\alpha_\mathrm{d} = \pi/2,$ $\kappa_\mathrm{w}=1$ without the kernel term, with the kernel but without the bell curve weight, and with the kernel using the bell curve.
    It is shown that the penalization of the attitude deviation from the kernel space is crucial to stabilize the system.
    Furthermore, note that the inclusion of the bell curve yields a faster and more efficient response, as it establishes an effective region where the kernel deviation penalty is activated.
    %
    %
    %
    %
    %
    }
    \label{fig:example1_case1}
\end{figure}

 Consider the UAV-object planar system \eqref{eq:UAV_sys_canonical1}, \eqref{eq:UAV_sys_canonical2} depicted in Figure \ref{fig:UAV_manipulating_object}.
Let $m_\mathrm{u} = 100$ g, $I_\mathrm{u} = 1.014$ g$^2$m$^2,$ $m_\mathrm{o} = 30$ g, $I_\mathrm{o} = 2$ kg$^2$m$^2,$ and $L=1.25$ m.
The constraint on the thrust is $0$ N $\le u_1 \le 5$ N and the constraint on the torque is $|u_2| \le 0.2$ Nm.
Note that these parameters come from the real experimental testbed used in \cite{NGUYEN2019384}.
The control objective is to stabilize the object to the desired angle $\alpha_\mathrm{d} \in [0, \pi].$
The KPCA uses $\mathcal{K}(\cdot) = [x_{1,1} \; 0]^\mathrm{T},$ $Q = \mathrm{diag}\{3, \allowbreak 1,\allowbreak 2,\allowbreak 5\},$ $R=\mathrm{diag}\{1,\allowbreak 0.01\},$ $T_\mathrm{s} = 0.1$ s, and $N = 15.$
Note that, in this example, since $x_\mathrm{d} = [\alpha_{\mathrm{d}} \; 0 \; \beta_\mathrm{d} \; 0]^\mathrm{T},$ only the first component of $x_\mathrm{2,d}$ in \eqref{eq:NLP} is the optimization variable.
Furthermore, we do not consider the equality constraint $\Pi(\cdot)$ in \eqref{eq:NLP}.
%
%
The maximum iteration number of IPOPT is 40 and the minimum barrier parameter is 0.1.
The nonlinear continuous controller (NCC) is given by \eqref{eq:control_law1}-\eqref{eq:control_law4} and uses
$k_{\mathrm{p},\alpha} = 4,$
$k_{\mathrm{d},\alpha} = 2.5,$
$k_{\mathrm{p},\beta}= 3\cdot 10^{-5},$ $k_{\mathrm{d},\beta}= 10^{-5},$ 
and $\varepsilon = 0.4.$
%
%
Figure \ref{fig:example1_case1} compares the responses of KPCA and NCC  for $x_0 = [0 \; 0 \; \pi/6 \; 0]^\mathrm{T}$ using $\alpha_\mathrm{d} = \pi/2,$ $\kappa_\mathrm{w} = 1,$ without the kernel, with the kernel without the bell curve, and with the kernel using the bell curve.
%
%

\subsection{Example 2: UAV manipulating an object in 3D}

\begin{figure}
    \centering
    \includegraphics[width=.4\columnwidth]{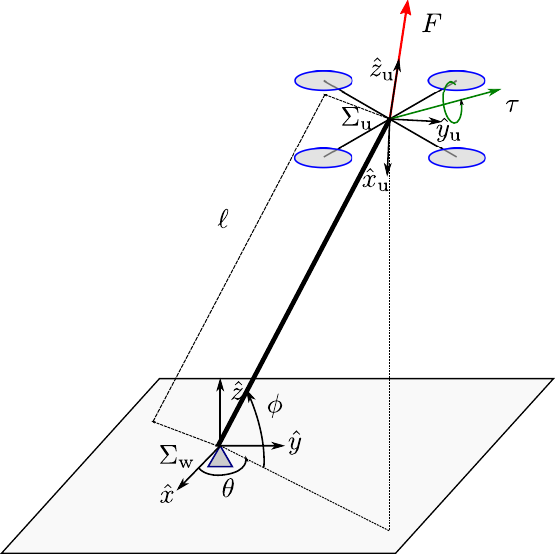}
    \caption{UAV manipulating an object in three dimensions.
    }
    \label{fig:UAV3d}
\end{figure}

Consider the UAV manipulating an object in three dimensions as depicted in Figure \ref{fig:UAV3d}.
Assume that the UAV center of mass coincides with the joint and the object center of mass is halfway the distance between the ground joint and the UAV joint.
Using the Euler-Lagrange method, the three-dimensional UAV-object system is derived as
\begin{align}
    \ddot{q}  & = M(q)^{-1} [\mathcal{F}(q, \mathbf{q}, T) - C(q,\dot{q})\dot{q} - G(q)], \label{eq:UAV3D_1} \\
    \dot{\mathbf{q}} & = \frac{1}{2} E(\mathbf{q})\omega, \quad\quad
    \dot{\omega} = I_\mathrm{u}^{-1}(-\hat{\omega}I_\mathrm{u}\omega + \tau), \label{eq:UAV3D_3}
\end{align}
where
$I_3\in\mathbb{R}^{3\times3}$ is the 3-by-3 identity matrix,
$\forall x\triangleq[x_1 \; x_2 \; x_3]^\mathrm{T}\in\mathbb{R}^3, \hat{x}:\mathbb{R}^3\to so(3)$ is the skew operator and is defined by
$\hat{x} \triangleq \left[
    \begin{smallmatrix}
    0 & -x_3 & x_2 \\
    x_3 & 0 & -x_1 \\
    -x_2 & x_1 & 0
    \end{smallmatrix}\right],$
$q \triangleq [\phi \; \theta]^\mathrm{T}$ is the generalized coordinate of the object, 
$\phi\in[0,\pi]$ is the polar angle,
$\theta\in[-\pi, \pi)$ is the azimuthal angle,
$\mathbf{q} \triangleq [q_0 \; q_\mathrm{v}^\mathrm{T}]^\mathrm{T}\in\mathbb{H}$ is the quaternion representing the rotation from the world frame $\Sigma_w$ to the UAV frame $\Sigma_u,$ $q_0\in\mathbb{R}$ is the real part, $q_\mathrm{v}\in\mathbb{R}^{3\times1}$ is the imaginary part,
$E(\mathbf{q}) \triangleq \left[\begin{smallmatrix}-q_\mathrm{v} & q_0 I_3 + \hat{q}_\mathrm{v}\end{smallmatrix}\right]^\mathrm{T},$
$\omega\in\mathbb{R}^{3\times1}$ is the UAV angular velocity,
$I_\mathrm{u}\in\mathbb{R}^{3\times3}_{>0}$ is the positive-definite UAV inertia matrix,
$\tau\in\mathbb{R}^{3\times1}$ is the torque applied to the UAV, 
$
    M(q) \triangleq
    \left[
    \begin{smallmatrix}
    I_{\mathrm{o},y} + \frac{\tilde{m}\ell^2}{2}(1-\cos(2\phi)) & 0 \\
    0 & I_{\mathrm{o},z} + \tilde{m}\ell^2\cos^2\phi
    \end{smallmatrix}
    \right],
    C(q,\dot{q}) \triangleq 
    \left[
    \begin{smallmatrix}
    \frac{\tilde{m}\ell^2}{2}\sin(2\phi)\dot\phi & 
    \frac{\tilde{m}\ell^2}{2}\sin(2\phi)\dot\theta \\
    -\frac{\tilde{m}\ell^2}{2} \sin(2\phi)\dot{\theta} & -\frac{\tilde{m}\ell^2}{2} \sin(2\phi)\dot{\phi}
    \end{smallmatrix}
    \right],
    G(q) \triangleq
    \left[
    \begin{smallmatrix}
    g\ell\cos\phi(\frac{m_\mathrm{o}}{2} + m_\mathrm{u}) \\
    0
    \end{smallmatrix}
    \right],
$
$\tilde{m} \triangleq \frac{m_\mathrm{o}}{4} + m_\mathrm{u},$
$I_\mathrm{o}\triangleq \mathrm{diag}\{I_{\mathrm{o},x}, I_{\mathrm{o},y}, I_{\mathrm{o},z}\}\in\mathbb{R}^{3\times3}_{>0}$ is the positive-definite object inertia matrix,
$\ell\in\mathbb{R}_{>0}$ is the distance between the ground joint and the UAV joint,
$m_\mathrm{o}\in\mathbb{R}_{>0}$ is the object mass,
$m_\mathrm{u}\in\mathbb{R}_{>0}$ is the UAV mass,
and $g\in\mathbb{R}_{>0}$ is the gravity acceleration.

Let $\mathcal{R}(\mathbf{q})\in SO(3)$ be the rotation matrix based on $\mathbf{q},$ which can be obtained by the Euler-Rodrigues formula \cite{DAI2015144}.
The generalized force is given by $\mathcal{F}(q, \mathbf{q}, T) =
    \mathcal{F}(q, \mathbf{q}, T) = \mathcal{P}_{\theta\phi} F(\mathbf{q}, T) \in\mathbb{R}^{2\times1},$
where $F(\cdot) \triangleq \mathcal{R}(\mathbf{q})[0 \; 0 \; T]^\mathrm{T}\in\mathbb{R}^{3\times1}$ is the three-dimension force generated by the UAV, $T\in\mathbb{R}_{\ge0}$ is the thrust, and
$
    \mathcal{P}_{\theta\phi} \triangleq \left[\begin{smallmatrix}\hat{\phi} \\ \hat{\theta}\end{smallmatrix}\right] = \left[
    \begin{smallmatrix}
    -\cos\theta\sin\phi & -\sin\phi\sin\theta & \cos\phi \\
    -\sin\theta & \cos\theta & 0
    \end{smallmatrix}
    \right]
$
is the projection matrix from the Euclidian coordinates $(\hat{x},\hat{y},\hat{z})$ to the polar coordinates $(\hat{\phi},\hat{\theta}).$

Let $x_1 = [q^\mathrm{T} \; \dot{q}^\mathrm{T}]^\mathrm{T},$ $x_2 = [\mathbf{q}^\mathrm{T} \; \omega^\mathrm{T}]^\mathrm{T},$ $u_1 = T,$ $u_2 = \tau,$ $x_{1,1} \triangleq q,$ $x_{1,2} \triangleq \dot{q},$ $x_{2,1} \triangleq \mathbf{q},$ and $x_{2,2} \triangleq \omega.$
We can rewrite \eqref{eq:UAV3D_1}-\eqref{eq:UAV3D_3} in the canonical form \eqref{eq:S1}, \eqref{eq:S2} as
\begin{align}
    \dot{x}_1 &= M(x_1)^{-1} [\mathcal{F}(x_1,x_2,u_1) - C(x_1)x_{2,1} - G(x_1)] \nonumber \\
    &= f(x_1, x_2, u_1), \label{eq:3d_can1} \\
    \dot{x}_2 &= \left[
    \begin{smallmatrix}
    \frac{1}{2} E(x_2)x_{2,2} \\
    I_\mathrm{u}^{-1}(-\hat{x}_{2,2}I_\mathrm{u}x_{2,2} + u_2)
    \end{smallmatrix}
    \right] = g(x_2, u_2). \label{eq:3d_can2}
\end{align}

According to \eqref{eq:Psi}, the effective control is given by
$
    \tilde{u}_1 = \mathcal{F}(x_1, x_2, u_1).
$
Note that the system is overactuated since $\dim(u_1) + \dim(x_2) = 8 > \dim(\tilde{u}_1) = 2,$ which complies with Assumption \ref{ass:overactuated}.
Using \eqref{eq:kernel}, the kernel of $\Psi(\cdot)$ is
\begin{align}\label{eq:kernelUAV3D}
    \ker(\Psi(\cdot)) = \{(x_1,x_2,u_1):\mathcal{F}(x_1,x_2,u_1)=0\}.
\end{align}
%
%
First, note that $\mathcal{F}(\cdot)\ne0$ if and only if $F(\cdot)$ has a nonzero component along $\hat\theta$ or $\hat\phi.$
Next, let $\hat{r}\in\mathbb{R}^3$ be the radial unit vector, which is perpendicular to $\hat{\theta}$ and $\hat{\phi}.$
Then, to produce $\mathcal{F}(\cdot)=0$ for any $F(\cdot)\ne0,$ it is necessary that the UAV is oriented such that $F(\cdot)$ is aligned with $\hat{r}.$
Note that, however, for $F(\cdot)=0,$ any UAV attitude necessarily produces $\mathcal{F}(\cdot)=0.$

Specifically, for all $x_1\in\mathbb{R}^{n_1}$ and for all $u_1^*\in\mathbb{R}^{m_1},$ there exists a class $C^0$ function $\mathcal{K}(\cdot)$ such that $x_2^* = \mathcal{K}(x_1) \Rightarrow \Psi(x_1,x_2^*,u_1^*) = 0,$ where $\mathcal{K}(\cdot)\in\mathcal{S}_\mathcal{K},$ and $\forall\omega_0\in\mathbb{R}^3,$ $\forall n\in\mathbb{Z},$ $\forall\psi\in\mathbb{R},$
$
\mathcal{S}_\mathcal{K} \triangleq \left\{\mathcal{K}_n:\mathbb{R}^4\to\mathbb{R}^7: \mathcal{K}_n(x_1) = [\mathbf{q}_{0,n}^\mathrm{T} \; \; \omega_0^\mathrm{T}]^\mathrm{T}\right\},
$
where
$\mathbf{q}_{n,0} \triangleq \left[
    \begin{smallmatrix}
    c_{\hat{\psi}} c_{\hat{\theta}} c_{\hat{\phi}_n} - c_{\hat{\phi}_n} s_{\hat{\psi}} s_{\hat{\theta}} \\
    c_{\hat{\theta}} s_{\hat{\psi}} s_{\hat{\phi}_n} - c_{\hat{\psi}} s_{\hat{\theta}} s_{\hat{\phi}_n} \\
    -c_{\hat{\psi}} c_{\hat{\theta}} s_{\hat{\phi}_n} - s_{\hat{\psi}} s_{\hat{\theta}} s_{\hat{\phi}_n} \\
    c_{\hat{\psi}} c_{\hat{\phi}_n} s_{\hat{\theta}} + c_{\hat{\theta}} c_{\hat{\phi}_n} s_{\hat{\psi}}
    \end{smallmatrix}
    \right],
$
$c_\cdot \triangleq \cos(\cdot),$ and $s_\cdot \triangleq \sin(\cdot),$ where the indices of $c_\cdot$ and $s_\cdot$ are defined by $\hat{\psi}\triangleq \frac{\psi}{2},$ $\hat{\theta}\triangleq \frac{\theta}{2},$ and $\hat{\phi}_n\triangleq \frac{\phi}{2} - \frac{\pi}{4} + n\frac{\pi}{2}.$
Note that this fact complies with Assumption \ref{ass:kernel}.

Note that the system \eqref{eq:3d_can1}, \eqref{eq:3d_can2} linearized around the equilibrium point $(\bar{x}, \bar{u}) = ([\frac{\pi}{2} + m\pi \allowbreak \; \bar{\theta} \allowbreak \; 0 \allowbreak \; 0 \allowbreak \; \mathbf{q}_{n,0} \allowbreak \; 0 \allowbreak \; 0 \allowbreak \; 0]^\mathrm{T}, [0 \; 0 \; 0 \; 0]^\mathrm{T})$ for all $n,m\in\mathbb{Z}$ and for all $\bar\theta\in[-\pi,\pi)$ is uncontrollable.
This fact complies with Assumption \ref{ass:controllability}.

Next, let $p\in\mathbb{R}^{3\times1}$ be the UAV position, where
\begin{align} \label{eq:p}
    p &= 
    \left[
    \begin{matrix}
    \ell \cos\phi \cos\theta &
    \ell \cos\phi \sin\theta &
    \ell \sin\phi
    \end{matrix}
    \right]^\mathrm{T},
\end{align}
and denote the desired UAV position by $p_\mathrm{d}\in\mathbb{R}^{3\times1},$ which is computed by \eqref{eq:p} with $\theta = \theta_\mathrm{d}$ and $\phi = \phi_\mathrm{d},$ where $\theta_\mathrm{d}\in[-\pi,\pi)$ and $\phi_\mathrm{d}\in[0,\pi]$ are the desired azimuthal and polar angles, respectively.

Inspired by \cite{doi:10.2514/1.G004356}, we propose the modified, nonlinear continuous geodesic-based control law
\begin{align}\label{eq:Td}
    T \mathcal{R}_\mathrm{d}(\mathbf{q}_\mathrm{d}) \hat{z} & = T_t\hat{t} + T_\phi \hat{\phi} + T_r \hat{r},
\end{align}
where $T \mathcal{R}_\mathrm{d}(\mathbf{q}_\mathrm{d}) \hat{z} \triangleq [T_{\mathrm{d},x} \; T_{\mathrm{d},y} \; T_{\mathrm{d},z}]^\mathrm{T}\in\mathbb{R}^{3\times1}$ is the desired thrust in three dimensions,
$T = \sqrt{T_{\mathrm{d},x}^2 + T_{\mathrm{d},y}^2 + T_{\mathrm{d},z}^2},$
$\mathcal{R}_\mathrm{d}(\mathbf{q}_\mathrm{d})\in SO(3)$ is the desired rotation matrix based on the desired quaternion $\mathbf{q}_\mathrm{d}\in\mathbb{H},$
$\hat{t}\in\mathbb{R}^{3\times1}$ is the geodesic direction,
$T_\phi\triangleq G(q)$ is the gravity compensation,
%
%
$T_r \in \mathbb{R}_{>0}$ is arbitrarily small,
$\hat{r}\triangleq \frac{p}{\ell},$
$
    T_t\hat{t} \triangleq \mathrm{dist}(p, p_\mathrm{d}) k_{p,t}\hat{t}
    - k_{\mathrm{d},t} \dot{p},
    \mathrm{dist}(p,p_\mathrm{d}) \triangleq \ell\arccos\left(\left\langle \frac{p}{\ell}, \frac{p_\mathrm{d}}{\ell} \right\rangle\right),
$
$\hat{t} \triangleq \frac{(p\times p_\mathrm{d})\times p}{\max\{\|(p\times p_\mathrm{d})\times p\|_2,\varepsilon\}},$
$\langle \cdot, \cdot \rangle : \mathbb{R}^3\times\mathbb{R}^3\to\mathbb{R}_{\ge0}$ is the dot product,
$\cdot \times \cdot : \mathbb{R}^3\times\mathbb{R}^3\to\mathbb{R}_3$ is the cross product,
$k_{\mathrm{p},t}\in\mathbb{R}_{>0}$ and $k_{\mathrm{d},t}\in\mathbb{R}_{>0}$ are the geodesic control gains,
and
$\varepsilon\in\mathbb{R}_{>0}$ is an arbitrarily small scalar.
%
%
%
The desired quaternion $\mathbf{q}_\mathrm{d}$ is
\begin{align}\label{eq:qd}
    \mathbf{q}_\mathrm{d} = \left[
    \begin{smallmatrix}
    q_{\zeta,0} & -q_{\zeta,v}^\mathrm{T} \\
    q_{\zeta,v} & q_{\zeta,0}I_3 + \hat{q}_{\zeta,v}
    \end{smallmatrix}
    \right]
    \left[
    \begin{smallmatrix}
    q_{\psi,0} \\
    q_{\psi,v}
    \end{smallmatrix}
    \right],
\end{align}
where $\psi\in[-\pi,\pi)$ is an arbitrary desired yaw angle\footnote{In this paper, $\psi=0$ for simplicity.}, $q_{\psi,0} \triangleq \cos(\psi/2),$ $q_{\psi,v} \triangleq [0 \; 0 \; \sin(\psi/2)]^\mathrm{T},$
$
    q_{\zeta, 0} \triangleq \cos\frac{\zeta_\mathrm{d}}{2},
    q_{\zeta, v} \triangleq \frac{\sin(\zeta_\mathrm{d}/2)}{\sqrt{T_{\mathrm{d},x}^2 + T_{\mathrm{d},y}^2}} \left[
    \begin{smallmatrix}
    -T_{\mathrm{d},y} \\
    T_{\mathrm{d},x} \\
    0
    \end{smallmatrix}
    \right],
    \zeta_\mathrm{d} \triangleq \arctan2\left(\sqrt{T_{\mathrm{d}, x}^2 + T_{\mathrm{d}, y}^2}, T_{\mathrm{d},z}\right),
$
and $\arctan2(y,x): \mathbb{R}\times\mathbb{R}\to\mathbb{R}.$
Note that, for $T_{\mathrm{d},x} = T_{\mathrm{d},y} = 0,$ we have $q_{\zeta,0} = 1$ and $q_{\zeta,v} = [0 \; 0 \; 0]^\mathrm{T}.$

Next, we control the UAV attitude through the nonlinear continuous quaternion-based control law
\begin{align}\label{eq:tau_3d}
    \tau &= k_{\mathrm{p}, \mathbf{q}} \tilde{\mathbf{q}}_v - k_{\mathrm{d}, \mathbf{q}} \omega,
\end{align}
where $k_{\mathrm{p},\mathbf{q}} \in \mathbb{R}_{>0}$ and $k_{\mathrm{d},\mathbf{q}} \in \mathbb{R}_{>0}$ are the attitude control gains, and $\tilde{q}_v$ is the imaginary part of the quaternion $\tilde{\mathbf{q}}\in\mathbb{H},$ which is obtained by the inverse Euler-Rodrigues formula of
$
    \tilde{\mathcal{R}}(\tilde{\mathbf{q}}) = \mathcal{R}^\mathrm{T}(\mathbf{q})\mathcal{R}_\mathrm{d}(\mathbf{q}_\mathrm{d}).
$

Note that the closed-loop system \eqref{eq:3d_can1}, \eqref{eq:3d_can2} controlled by \eqref{eq:Td} can be rewritten in the canonical form \eqref{eq:S1_controlled}, \eqref{eq:S2_controlled} using $\mathcal{R}(\cdot) = \mathcal{R}_\mathrm{d}(\cdot) + \mathcal{R}_\mathrm{d}(\cdot)(\tilde{\mathcal{R}}^\mathrm{T}(\cdot) - I_3).$
%
Moreover, note that $T_r > 0$ ensures that \eqref{eq:qd} is Lipschitz continuous.
%
%
Hence, since Assumptions \ref{ass:exp_stability} and \ref{ass:iss} \cite{NGUYEN2021109586} are satisfied and the allocated mapping is Lipschitz continuous, according to Theorem \ref{th:UAS}, the closed-loop system is UAS.
Furthermore, note that the steps to derive the solution from the generalized inverse are similar to the previous case (see Section \ref{sec:case_study}).
Because the mapping is also discontinuous, it follows directly from Theorem \ref{th:UAS} that the closed-loop system using the generalized inverse is unstable.


Let $m_\mathrm{u} = 150$ g, $I_\mathrm{u} = \mathrm{diag}\{0.005, 0.005, 0.01\}$ kg$^2$m$^2,$ $m_\mathrm{o} = 100$ g, $\ell = 3$ m, and $I_\mathrm{o} = \mathrm{diag}\{0.01, 2, 2\}$ kg$^2$m$^2$.
The constraint on the thrust is 0 N $\leq u_1 \leq$ 7 N and the constraint on the torque is $|u_i| < 0.5$ Nm $(i=2,3,4).$ 
Let $x\triangleq [\phi \allowbreak \; \theta \allowbreak \; \dot\phi \allowbreak \; \dot\theta \allowbreak \; \mathbf{q}^\mathrm{T} \allowbreak \; \omega^\mathrm{T}]^\mathrm{T}.$
The control objective is to stabilize the object to the desired azimuthal and polar angles $\theta_\mathrm{d}\in[-\pi,\pi)$ and $\phi_\mathrm{d}\in[0, \pi],$ respectively.

\begin{figure}
    \centering
    \includegraphics[width=.45\columnwidth, trim={0.9cm 1.8cm 0.9cm 1.3cm}]{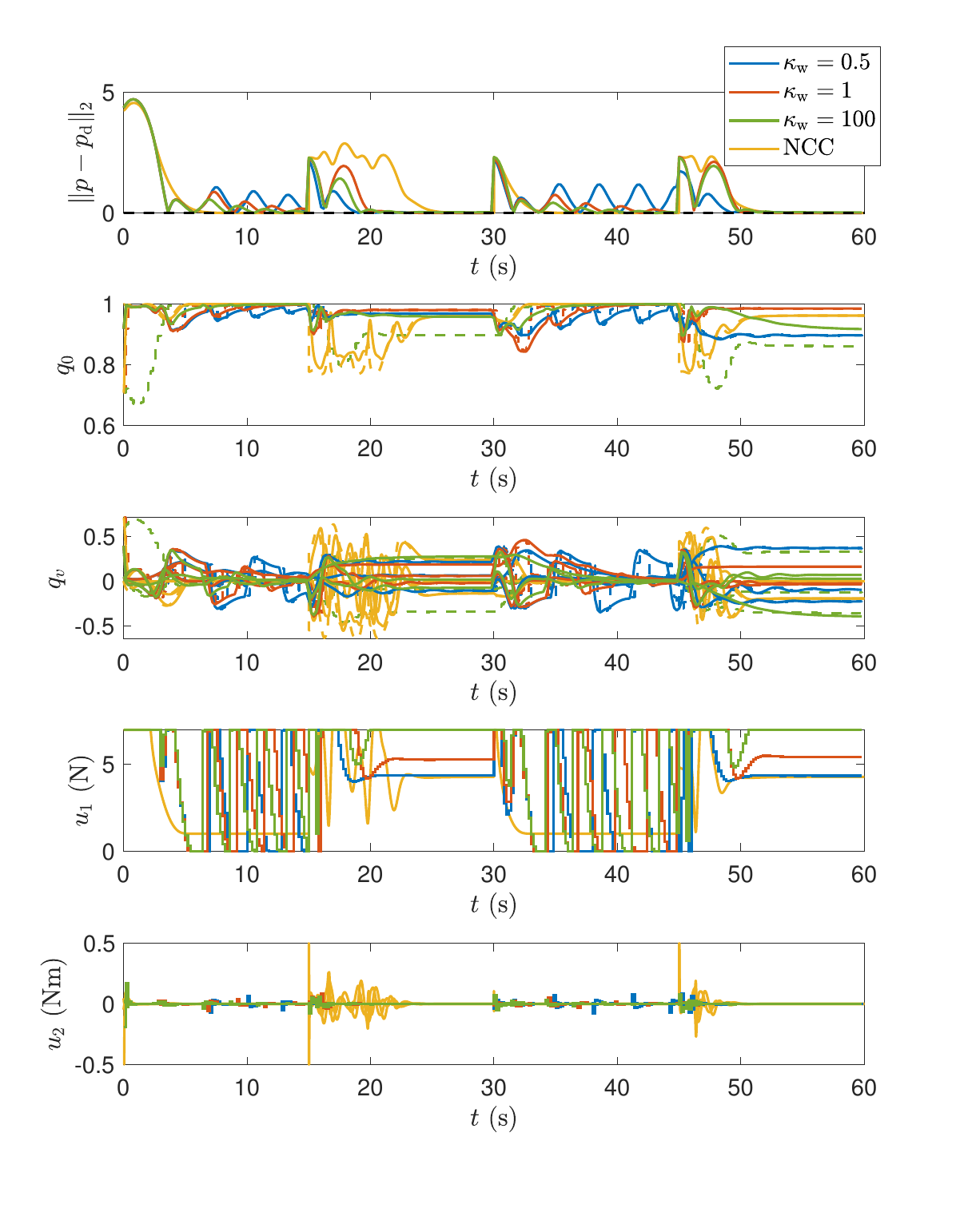}
    \caption{\textbf{Example 2.}
    Multiple-step command following using $\kappa_\mathrm{p}=10$ and various values of $\kappa_\mathrm{w}.$
    We see that all responses converge to the references.
    However, note that, as $\kappa_\mathrm{w}$ decreases, the oscillations become more prominent.
    %
    %
    Moreover, note that the NCC thrust has the least control-effort trajectory but is the greediest in terms of the torque efforts.
    As $\kappa_\mathrm{w}\to 0,$ the thrust control effort from KPCA decreases for the second and fourth step commands, but the thrust bursts several times for the first and third step commands.
    }
    \label{fig:3d_case2}
\end{figure}

The KPCA uses \eqref{eq:kernelUAV3D} with $n=0,$ $\psi=0,$ and $\omega_0 = [0 \; 0 \; 0]^\mathrm{T} \equiv 0_{3\times1},$ that is,
$
    \mathcal{K}(x_1) = \left[
    \begin{smallmatrix}
    \cos\frac{\theta}{2}\cos\left(\frac{\phi}{2} - \frac{\pi}{4}\right) \\
    -\sin\frac{\theta}{2}\sin\left(\frac{\phi}{2} - \frac{\pi}{4}\right) \\
    -\cos\frac{\theta}{2}\sin\left(\frac{\phi}{2} - \frac{\pi}{4}\right) \\
    \sin\frac{\theta}{2}\cos\left(\frac{\phi}{2} - \frac{\pi}{4}\right) \\
    0_{3\times1}
    \end{smallmatrix}
    \right],
$
$Q = \mathrm{diag}\{20, \allowbreak 20, \allowbreak 0.5, \allowbreak 0.5, \allowbreak 1, \allowbreak 1, \allowbreak 1, \allowbreak 1, \allowbreak 0.5, \allowbreak 0.5, \allowbreak 0.5\},$ $R = \mathrm{diag}\{1, \allowbreak 0.01, \allowbreak 0.01, \allowbreak 0.01\},$ $T_\mathrm{s} = 0.2$ s, and $N=5.$
Furthermore, $\Pi(\cdot) \equiv \|\mathbf{q}_\mathrm{d}\| - 1 = 0,$ and the first four components of $x_{2,\mathrm{d}}$ (that is, the desired quaternion $\mathbf{q}_\mathrm{d}$) are the optimization variables in \eqref{eq:NLP}.
The maximum iteration number of IPOPT is 50 and the minimum barrier parameter is 0.1.
The NCC is given by \eqref{eq:Td} and uses the control parameters $k_{\mathrm{p}, t} = 2,$ $k_{\mathrm{d}, t} = 3,$ $k_{\mathrm{p}, \mathbf{q}} = 2,$ $k_{\mathrm{d}, \mathbf{q}} = 0.2,$ $T_r = 1$ N, and $\varepsilon = 0.01.$
Figure \ref{fig:3d_case2} compares the tracking responses of KPCA and NCC for $x_0 = [0 \allowbreak \; 0 \allowbreak \; 0 \allowbreak \; 0 \allowbreak \; \frac{\sqrt{2}}{2} \allowbreak \; \frac{\sqrt{2}}{2} \allowbreak \; 0 \allowbreak \; 0 \allowbreak \; 0 \allowbreak \; 0 \allowbreak \; 0]^\mathrm{T}$ using $\kappa_\mathrm{p}=5$ and various values of $\kappa_\mathrm{w},$ where the step-reference trajectory $q_\mathrm{d} \triangleq [\phi_\mathrm{d} \; \theta_\mathrm{d}]^\mathrm{T}$ is given by $[\frac{\pi}{2} \; 0]^\mathrm{T}$ for $0 \; \text{s} \le t < 15 \; \text{s},$ $[\frac{5\pi}{4} \; \frac{\pi}{6}]^\mathrm{T}$ for $15 \; \text{s} \le t < 30 \; \text{s}, $ $[\frac{\pi}{2} \; 0]^\mathrm{T}$ for $30 \; \text{s} \le t < 45 \; \text{s},$ and $[\frac{\pi}{4} \; \frac{-\pi}{4}]^\mathrm{T}$ for $45 \; \text{s} \le t.$
%
%

\subsection{Example 3: Control of a surface vessel actuated by two azimuthal thrusters}

Interestingly, this example shows that, even if we do not know the analytic solution a-priori, it is possible to directly apply KPCA to control the system.
\begin{figure}
    \centering
    \includegraphics[width=.4\columnwidth]{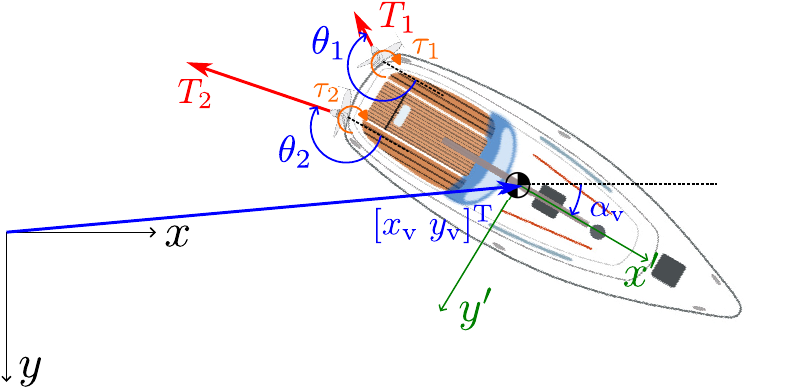}
    \caption{Surface vessel actuated by two azimuthal propellers.}
    \label{fig:surface_vessel}
\end{figure}
Consider the planar control of a surface vessel actuated by two azimuthal thrusters (Figure \ref{fig:surface_vessel}). This example is inspired by \cite{mukwege4457173continuity} and extends the model to the full 3-DOF dynamics.
Let $x_{\mathrm{v}},y_{\mathrm{v}}\in\mathbb{R}$ be the $x$- and $y$-cartesian coordinates of the center of mass of the surface vessel, respectively, and let $\alpha_{\mathrm{v}}\in(-\pi,\pi]$ be the heading angle of the surface vessel, which is the angle between the $x$-axis of the inertial frame and the heading direction of the surface vessel.
Furthermore, let $\theta_1, \theta_2\in[0,2\pi)$ be the angle between the heading direction and the thrust direction of the left and the right propellers, respectively.
Denote by $m_\mathrm{v}\in\mathbb{R}_{>0}$ and $\mathcal{I}_\mathrm{v}\in\mathbb{R}_{>0}$ the mass and the inertia of the surface vessel, respectively, and by $\mathcal{I}_\mathrm{p} \in\mathbb{R}_{>0}$ the inertia of a propeller.
For $x_\mathrm{v} = y_\mathrm{v} = \alpha_\mathrm{v} = 0,$ the left and the right propellers are symmetrically positioned at $[\ell_{x} \; \ell_{y}]^\mathrm{T}$ and $[\ell_{x} \; -\ell_{y}]^\mathrm{T}$ from the origin, respectively, where $\ell_{x}\in\mathbb{R}$ and $\ell_{y}\in\mathbb{R}_{\ge0}.$
The inputs are the left and the right torques $\tau_1,\tau_2\in\mathbb{R},$ respectively, and the left and right thrusts $T_1, T_2\in\mathbb{R}_{\ge0},$ respectively.

The equations of motion, in the canonical form \eqref{eq:S1}, \eqref{eq:S2}, are given by
\begin{align}
    \dot{x}_1 \hspace{-0.2em}= \hspace{-0.2em} \left[
    \begin{smallmatrix}
        \left[\begin{smallmatrix}0_{3\times3} & I_3\end{smallmatrix}\right] x_1 \\
        m_{\mathrm{v}}^{-1} [T_1\cos(\theta_1 + \alpha_{\mathrm{v}}) + T_2\cos(\theta_2 + \alpha_{\mathrm{v}})] \\
        m_{\mathrm{v}}^{-1} [T_1\sin(\theta_1 + \alpha_{\mathrm{v}}) + T_2\sin(\theta_2 + \alpha_{\mathrm{v}})] \\
        \mathcal{I}_{\mathrm{v}}^{-1} [\ell_{y} (T_1 c_{\theta_1} - T_2 c_{\theta_2}) -\ell_{x}(T_1 s_{\theta_1} + T_2 s_{\theta_2})]
    \end{smallmatrix}
    \right] \hspace{-0.2em} = \hspace{-0.2em} f(\cdot), \label{eq:vessel1}
\end{align}
\begin{align}
    \dot{x}_2 &= \left[
    \begin{smallmatrix}
    \left[\begin{smallmatrix}
        0_{2\times2} & I_2
    \end{smallmatrix}\right] x_2 \\
    \mathcal{I}^{-1} \tau_1 \\
    \mathcal{I}^{-1} \tau_2
    \end{smallmatrix}
    \right] = g(x_2,u_2), \label{eq:vessel2}
\end{align}
where $x_1 = [x_{\mathrm{v}} \allowbreak \; y_{\mathrm{v}} \allowbreak \; \alpha_{\mathrm{v}} \allowbreak \; \dot{x}_{\mathrm{v}} \allowbreak \; \dot{y}_{\mathrm{v}} \allowbreak \; \dot{\alpha}_{\mathrm{v}}]^\mathrm{T},$
$x_2 = [\theta_1 \; \theta_2 \; \dot{\theta}_1 \; \dot{\theta}_2]^\mathrm{T},$
$u_1 = [T_1 \; T_2]^\mathrm{T},$
and
$u_2 = [\tau_1 \; \tau_2]^\mathrm{T}.$

\begin{figure}
    \centering \includegraphics[width=.5\columnwidth, trim={0.7cm 2cm 0.7cm 0.2cm}]{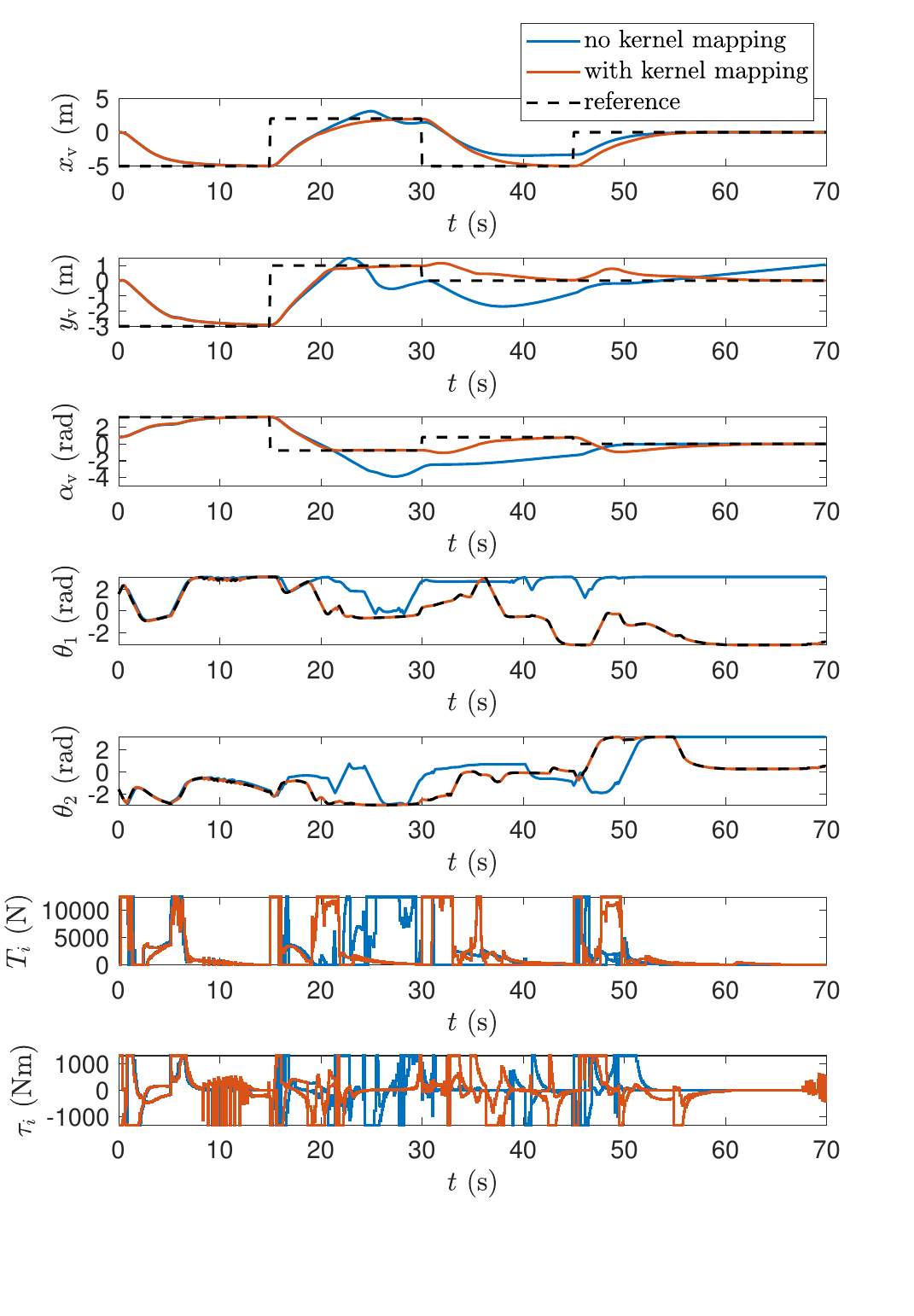}
    \caption{
    \textbf{Example 3.} Multiple-step command following with and without kernel mapping tracking for $i\in\{1,2\}.$ It is seen that, without the kernel mapping, the vessel is unable to track the second and third step commands. Furthermore, note that, without the kernel mapping, for the last step command, $y_\mathrm{v}$ drifts.
    }
    \label{fig:vessel_sim}
\end{figure}

The effective controller is given by 
\begin{align}
    \tilde{u}_1 &= \left[
\begin{smallmatrix}
    T[\cos(\theta_1+\alpha_{\mathrm{v}}) + \cos(\theta_2+\alpha_{\mathrm{v}})] \\
    T[\sin(\theta_1+\alpha_{\mathrm{v}}) + \sin(\theta_2+\alpha_{\mathrm{v}})] \\
    T [\ell_{y} (c_{\theta_1} - c_{\theta_2}) -\ell_{x}(s_{\theta_1} + s_{\theta_2})]
\end{smallmatrix}
\right] 
= \Psi(x_1,x_2,u_1).
\end{align}
Note that the system is overactuated since $\dim(u_1) + \dim(x_2) = 5 > \dim(\tilde{u}_1) = 3$ (Assumption \ref{ass:overactuated}).
Additionally, note that the kernel \eqref{eq:kernel} complies with Assumption \ref{ass:kernel} since, for all $x_1\in\mathbb{R}^{n_1}$ and for $u_1^*=[T^* \; T^*]^\mathrm{T},$ where $T^*\in\mathbb{R}_{\ge0},$ there exists a class $C^0$ function $\mathcal{K}(\cdot)\in\mathcal{S}_{\mathcal{K}},$ such that $x_2^* = \mathcal{K}(x_1) \Rightarrow \Psi(x_1,x_2^*,u_1^*) = 0.$ In particular, for all $\gamma_1,\gamma_2\in\mathbb{R}$ and for all $n\in\mathbb{Z},$ $
    \mathcal{S}_{\mathcal{K}} = \left\{
    \mathcal{K}_n(\cdot) =
    \left[
    \begin{matrix}
        \frac{\pi}{2}+n\pi &
        -\frac{\pi}{2}+n\pi &
        \gamma_1 &
        \gamma_2
    \end{matrix}
    \right]^\mathrm{T}
    \right\}.
$
Next, let $x = [x_1^\mathrm{T} \; x_2^\mathrm{T}]^\mathrm{T}$ and $u = [u_1^\mathrm{T} \; u_2^\mathrm{T}]^\mathrm{T}.$ The system \eqref{eq:vessel1}, \eqref{eq:vessel2} linearized around the equilibrium $\bar{x} = [\bar{x}_\mathrm{v} \allowbreak \; \bar{y}_\mathrm{v} \allowbreak \; \bar{\alpha}_\mathrm{v} \allowbreak \; 0 \allowbreak \; 0 \allowbreak \; 0 \allowbreak \; \pm\pi/2 \allowbreak \; \mp\pi/2 \allowbreak \; 0 \allowbreak \; 0]^\mathrm{T},$ $\bar{u} = [\bar{T} \; \bar{T} \; 0 \; 0]^\mathrm{T}$ is uncontrollable since
the controllability matrix has $\mathrm{rank}(\mathcal{C}) = 8 < 10,$ which complies with Assumption \ref{ass:controllability}.
The system parameters are inspired by the real Halcyon unmanned surface vehicle \cite{HEINS2017749}. Let $m_\mathrm{v} = 11000$ kg, $\mathcal{I}_\mathrm{v} = 36062$ kg$\cdot$m$^2$, $\mathcal{I}_\mathrm{p} = 700$ kg$\cdot$m$^2$, $\ell_x = -2.75$ m, and $\ell_y = 0.894$ m.
The control objective is to move the vessel to the desired position $(x_{\mathrm{d},\mathrm{v}},y_{\mathrm{d},\mathrm{v}})$ and desired heading angle $\alpha_{\mathrm{d},\mathrm{v}}$ considering the actuator saturations $0\le T_i \le 12.4$ kN and $|\tau_i| \le 1.325$ kN$\cdot$m and the propeller angle constraint $-\pi \le \theta_i \le \pi \; (i\in\{1,2\}).$
The KPCA uses $R=\mathrm{diag}\{10^{-10}, 10^{-10}, 10^{-4}, 10^{-4}]\},$ $T_\mathrm{s} = 0.1$ s, $N=5,$ and $\kappa_\mathrm{w} = 10^{-5}.$
Figure \ref{fig:vessel_sim} shows the responses of the controlled vessel for $\kappa_\mathrm{p} = 0$ with $Q=\mathrm{diag}\{20, \allowbreak 20,  \allowbreak 20,  \allowbreak 20,  \allowbreak 20,  \allowbreak 20,  \allowbreak 25,  \allowbreak 25,  \allowbreak 0,  \allowbreak 0\}$ (no mapping tracked) and $\kappa_\mathrm{p} = 10^{-6}$ with $Q=\mathrm{diag}\{20, 20, 20, 20, 20, 20, 50, 50, 25, 25\}$ (kernel mapping with tracking) using, for $\chi_\mathrm{d} \triangleq [x_{\mathrm{d},\mathrm{v}} \; y_{\mathrm{d},\mathrm{v}} \; \alpha_{\mathrm{d},\mathrm{v}}]^\mathrm{T},$ the step-command reference trajectory $[-5 \; -3 \; \pi]^\mathrm{T}$ for $0$ s $\le t < 15$ s, $[2 \; 1 \; -\pi/4]^\mathrm{T}$ for $15$ s $\le t < 30$ s, $[-5 \; 0 \; \pi/4]^\mathrm{T}$ for $30$ s $\le t < 45$ s, and $[0 \; 0 \; 0]^\mathrm{T}$ for $t \ge 45$ s.

\section{Conclusions}

This paper has presented a class of thrust vectoring systems, which are nonlinear, time-invariant, and overactuated. They consist of two subsystems and exhibit singular points, which are linearized uncontrollable.
To stabilize the system, we showed that we could use a two-level control-allocation scheme.
The control design was broken down into two separate control units for each subsystem.
To achieve the primary control objective, the two control units were connected through an allocated mapping, which was designed by properly formulating the secondary control objective.
Under adequate assumptions, we could prove that the system was asymptotically stable by using small gain arguments.
However, the bottleneck of this analytic solution was the design of a closed-form allocated mapping, which was in general nontrivial and not unique.
Additionally, this mapping must be Lipschitz continuous to guarantee asymptotic convergence.
Accordingly, we proposed a new kernel-based predictive control allocation, which took advantage of the kernel space to locally smooth out the generated allocated mapping.
We have demonstrated the effectiveness of the proposed scheme through three examples, which were the manipulation of an object through a UAV in two and three dimensions, and the control of a vessel actuated by two azimuthal thrusters.
Future works will aim at reducing the control efforts while minimizing the state deviation from the kernel space.
Note that the NLP formulation is not straightforward, as an undesired steady-state error can remain when trying to minimize the overall control efforts.
A more elaborated control strategy using terminal state constraints can be investigated in the future.


\bibliographystyle{plain}        
\bibliography{bibliography}           

\newpage

\parpic{\includegraphics[width=1in,height=1in,clip,keepaspectratio]{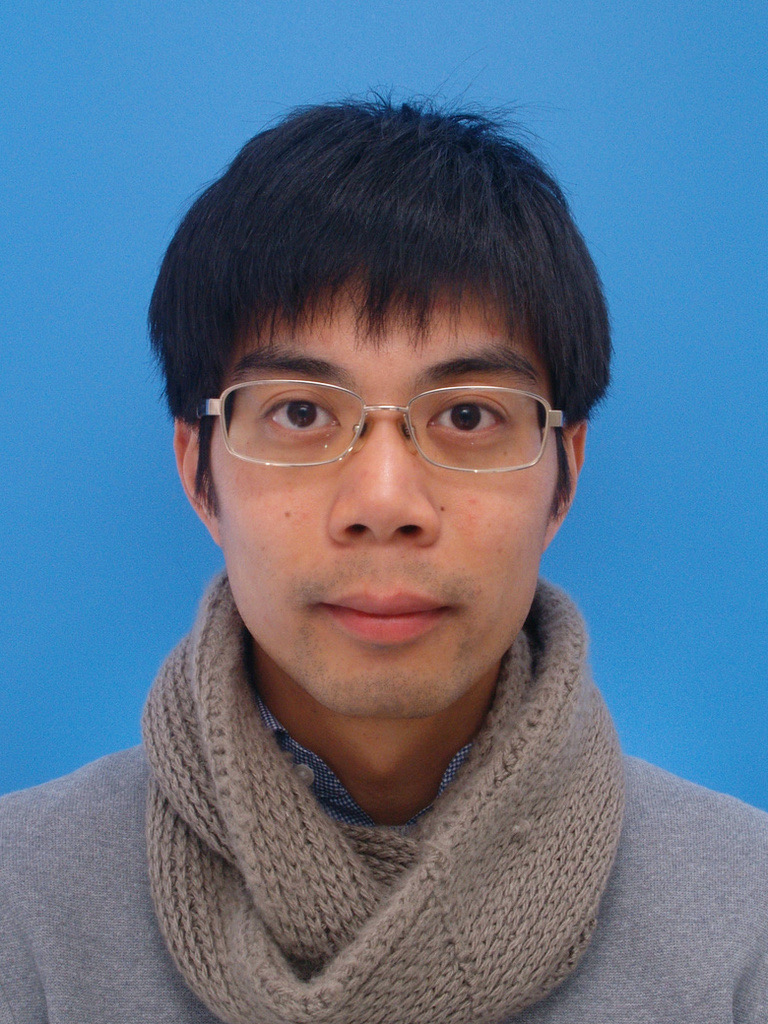}}
\textbf{Tam W. Nguyen} received the M.Sc. degree in mechatronics engineering and the Ph.D. degree in control systems from the Free University of Brussels, Belgium, in 2014 and 2018, respectively. He was a Postdoctoral Researcher at the Aerospace Engineering Department, University of Michigan, Ann Arbor, from 2019 to 2020. Since 2021, he has been an Assistant Professor with the Department of Electrical and Electronic Engineering, University of Toyama, Japan. His interests include nonlinear control, optimization control, and robotics.

\parpic{\includegraphics[width=1in,height=1in,clip,keepaspectratio]{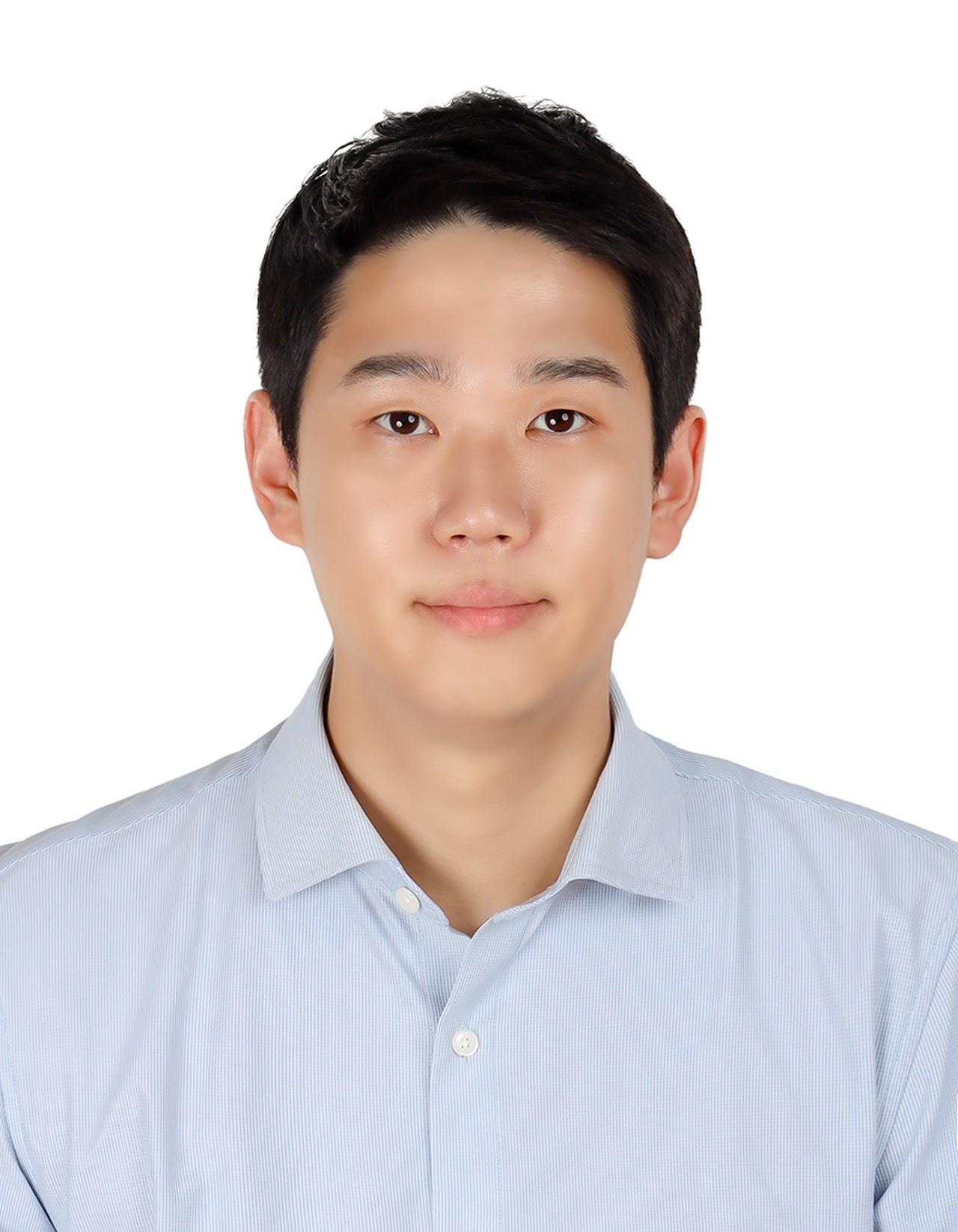}}
\textbf{Kyoungseok Han} received his B.S. degree in Civil Engineering with a minor in Mechanical Engineering from Hanyang University, Seoul, South Korea, in 2013, followed by M.S. and Ph.D. degrees in Mechanical Engineering from KAIST, Daejeon, South Korea, in 2015 and 2018, respectively. He is currently an Associate Professor in the Department of Automotive Engineering at Hanyang University, Seoul, South Korea. Prior to this, he worked as a Postdoctoral Research Fellow at the University of Michigan from 2018 to 2020 and as an Associate Professor at Kyungpook National University, Daegu, South Korea from March 2020 to August 2024. His research interests include autonomous vehicle modeling and control, energy-efficient control of electric vehicles, reinforcement learning, and optimal control theory and its applications.

\parpic{\includegraphics[width=1in,height=1in,clip,keepaspectratio]{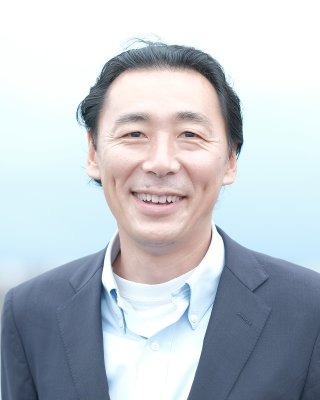}}
\textbf{Kenji Hirata} received the Ph.D. degree in information science from the Japan Advanced Institute of Science and Technology, Hokuriku, Japan. He is currently a Professor with the Faculty of Engineering, University of Toyama, Toyama, Japan. Prior to that appointment, he held faculty appointments at Osaka University, Suita, Japan, and Nagaoka University of Technology, Nagaoka, Japan. He was a Visiting Research Scholar with the University of California at Santa Barbara, Santa Barbara, CA, USA, from 2008 to 2009 and from 2012 to 2013. His research interests include the analysis and design of controls for systems with constraints, hybrid and switched dynamical systems, and distributed decision making in power demand/supply networks.

\end{document}